\documentclass[journal,twoside,final]{IEEEtran}

\usepackage{graphicx}
\usepackage{amsmath,bbm,epsfig,amssymb,amsfonts, amstext, verbatim,amsopn,cite,subfigure,multirow,multicol,lipsum}
\usepackage{balance}
\usepackage{url}
\usepackage{amsfonts}
\usepackage{epsfig}
\usepackage{epstopdf}
\usepackage{setspace}
\usepackage{stmaryrd}
\usepackage{psfrag}	
\usepackage{multirow}
\usepackage{float}
\usepackage[process=auto]{pstool}
\usepackage{etoolbox}
\usepackage{algorithm}
\usepackage{algorithmic}
\usepackage{hyperref}

\usepackage{pifont}
\allowdisplaybreaks

\newtheorem{theo}{Theorem}
\newtheorem{lem}{Lemma}
\newtheorem{remk}{Remark}

\newtheorem{corol}{Corollary}

\newtoggle{OneColumn}

\togglefalse{OneColumn}

\iftoggle{OneColumn}{%
\hoffset -4mm
\textwidth 17.9 cm
\textheight 23.7 cm 
}{%
\textwidth 18 cm
\textheight 23.8 cm 
}

\EndPreamble
\begin{document}

\title{Statistical Modeling of the FSO Fronthaul Channel \\ for UAV-based Communications}
\author{Marzieh Najafi, \textit{Student Member, IEEE,} Hedieh Ajam, \textit{Student Member, IEEE,} Vahid Jamali, \textit{Member, IEEE,}  Panagiotis D. Diamantoulakis, \textit{Senior Member, IEEE,}\\ George K. Karagiannidis, \textit{Fellow, IEEE,} and Robert Schober, \textit{Fellow, IEEE}\\  
\thanks{Marzieh Najafi, Hedieh Ajam, Vahid Jamali, Panagiotis D. Diamantoulakis, and Robert Schober are with the Institute for Digital Communications, Friedrich-Alexander-University Erlangen-Nuremberg (FAU), Germany (email:\{marzieh.najafi, hedieh.ajam, vahid.jamali, robert.schober\}@fau.de and padiaman@ieee.org).}

\thanks{George K. Karagiannidis is with the Aristotle University of Thessaloniki, Greece (email:geokarag@auth.gr).}

\thanks{This paper was presented in part at IEEE ICC 2018 \cite{ICC_2018}.}
}

\maketitle
\begin{abstract} 
In this paper, we investigate the statistics of the free space optics (FSO) communication channel between a hovering unmanned aerial vehicle (UAV) and a central unit. Two unique characteristics make UAV-based FSO systems significantly different from conventional FSO systems with static transceivers. First, for UAV-based FSO systems, the incident laser beam is not always orthogonal to the receiver lens plane. %
 Second, both position and orientation of the UAV fluctuate over time due to dynamic wind load, inherent random air fluctuations in the atmosphere around the UAV, and  internal vibrations of the UAV. On the contrary, for conventional FSO systems, the laser beam is always perpendicular to the receiver lens plane and the relative movement of the transceivers is limited.   
 In this paper, we develop a novel channel model for UAV-based FSO systems by quantifying the corresponding geometric and misalignment losses (GML), while taking into account the non-orthogonality of the laser beam and the random fluctuations of the position and orientation of the UAV. In particular, for diverse weather conditions, we propose  different fluctuation models for the position and orientation of the UAV and derive corresponding statistical models for the GML. We further analyze the performance of a UAV-based FSO link in terms of outage probability and ergodic rate and simplify the resulting analytical expressions for the high signal-to-noise ratio (SNR) regime. 
Finally, simulations validate the accuracy of the presented analysis and provide important insights for system design. For instance, we show that for a given variance of the fluctuations, the beam width should be properly adjusted to minimize the outage probability.
\end{abstract}


\section{Introduction}
Recently, there has been a growing interest in unmanned aerial vehicles (UAVs) for civil applications, such as delivering wireless access to remote regions or areas where a large number of users is \textit{temporarily} gathered, e.g., for a football match or a live concert, and permanent infrastructure is not available or is costly to deploy \cite{Alouini_Drone,Data_Collection_UAV_FSO}. 
In particular, UAVs may hover above the desired area and operate as mobile remote radio heads to assist the communication between the users and a central unit (CU) \cite{Alouini_Drone}. 

For these applications, free space optics (FSO) communication has been considered as a promising candidate for fronthauling of the data gathered by the UAVs to the CU \cite{Alouini_Drone,FSO_Survey_Murat,Optic_Tbit}. FSO systems offer the large bandwidth needed for data fronthauling, while FSO transceivers are relatively cheap compared to their radio frequency (RF) counterparts and easy to deploy \cite{Optic_Tbit,MyTCOM}. However,  the quality of the FSO link between a hovering UAV and a CU is negatively affected by variations (jitters) of the position and orientation of the UAV, which originate from several sources including dynamic wind load, inherent random air fluctuations in the atmosphere around the UAV, and internal vibrations of the UAV caused by the rotation of its propellers. These variations directly affect the performance of the tracking system, which is responsible for aligning the beam with the receiver lens at the CU \cite{Controller_FSO_Komaee,Controller_FSO_Arnon,Controller_FSO_Yuksel,Controller_FSO_Liu,Robust_laser_tracking}. 
Therefore, one important question is: \textit{How well (stable) does a UAV have to maintain its position and orientation in order to achieve a certain FSO link quality?} In this paper, we develop a mathematical framework for answering this question by statistically characterizing the geometric and misalignment losses (GML)\footnote{\label{Ftn:Loss} The receiver can only capture the fraction of power that falls on its lens. This phenomenon is known as \textit{geometric loss}. Moreover, misalignment of the center of the optical beam and the center of the receiver lens further increases the geometric loss. This phenomenon is known as \textit{misalignment loss} \cite{Steve_pointing_error}.} caused by the random fluctuations of the position and orientation of UAVs.

We note that even for conventional FSO systems with immobile transceivers fixed at building tops, random fluctuations of the positions of the transceivers occur due to building sway, which leads to random GML, known as pointing errors \cite{Steve_pointing_error,Alouini_Pointing,George_Pointing_error}. For this case, corresponding statistical models were derived in \cite{Steve_pointing_error} and \cite{Alouini_Pointing}. However, UAV-based FSO systems introduce the following new challenges: \textit{i)}~For conventional FSO links, it is typically assumed that the laser beam is orthogonal with respect to (w.r.t.) the receiver lens plane, as orthogonality maximizes the amount of laser power collected by the photo-detector (PD) located behind the lens \cite{Steve_pointing_error}. However, orthogonality may not hold for UAV-based FSO communication systems. For example, the position of a UAV may depend on the locations and traffic needs of the users, while the CU may not be able to adjust the orientation of the receiver lens due to limited mechanical capabilities. In addition, using one receiver lens and multiple PDs \cite{droz2014photodetector,hahn2010fiber}, the CU may receive data from several UAVs having different positions. Hence, it is not possible to orthogonally align the laser beams of all UAVs with the receiver lens plane. \textit{ii)} Unlike building sway, where the buildings exhibit limited movement due to wind loads and thermal expansion, for UAV-based FSO communication, both the position and orientation of the UAV may fluctuate over time and have to be modeled as random variables (RVs).

UAVs with FSO links have already been considered in the literature \cite{ICC_2018,Alouini_Drone,UAV_FSO,Data_Collection_UAV_FSO,Inter-UAV_FSO,Khalighi_FSO_UAV_ChannelModel,Dabiri_UAVChannelModel_2019,Mai_UAV2018,Mai_UAV_2019}. In particular, the authors of \cite{Alouini_Drone} discussed the advantages and challenges of FSO fronthauling in UAV-based networks. Moreover, the authors of \cite{UAV_FSO,Data_Collection_UAV_FSO,Inter-UAV_FSO} studied a system consisting of several UAVs that were connected with each other through FSO links. Specifically, the authors of \cite{UAV_FSO} presented a deterministic model for the geometric loss, assuming that the laser beam is always orthogonal to the receiver's lens plane. To the best of the authors' knowledge, a statistical model for the GML of a UAV-based FSO channel, which takes into account the fluctuations of the UAV's position and orientation as well as the non-orthogonality of the laser beam w.r.t. the receiver lens plane, has been reported first in the conference version of this paper \cite{ICC_2018}. Later on, the authors of  \cite{Khalighi_FSO_UAV_ChannelModel} derived a statistical model for the GML assuming random UAV positions and orientations, for the special case where the laser beam is orthogonal to the receiver lens plane and the variances of the fluctuations of the position (orientation) are identical for all directions. Moreover, in recent work, the authors of \cite{Dabiri_UAVChannelModel_2019,Mai_UAV2018,Mai_UAV_2019} developed statistical channel models for FSO links connecting different UAVs assuming  orthogonal beams. In particular, the authors of \cite{Dabiri_UAVChannelModel_2019} investigated the outage probability, which  was defined as the probability that the transmitting UAV falls out of the receiver's field-of-view (FoV) due to fluctuations of the orientation of the receiving UAV. Moreover, in \cite{Mai_UAV2018,Mai_UAV_2019}, adaptive beam control techniques were proposed to cope with the fluctuations of the positions and orientations of the transmitting and receiving UAVs. Unlike \cite{Alouini_Drone,UAV_FSO,Data_Collection_UAV_FSO,Inter-UAV_FSO,Khalighi_FSO_UAV_ChannelModel,Dabiri_UAVChannelModel_2019,Mai_UAV2018,Mai_UAV_2019}, we develop statistical channel models for UAV-based FSO communication systems that allow for non-orthogonal laser beams w.r.t. the receiver lens plane and take into account various models for the fluctuations of the position and orientation of the UAV. In particular, this paper makes the following contributions:

\begin{itemize}
\item We derive the GML for a given position and orientation of the UAV, which we refer to as \textit{conditional GML}. In particular, since obtaining a closed-form expression for the conditional GML is difficult, if not impossible, we first derive tight lower and upper bounds, and then provide a closed-form approximation based on these bounds. 
\item We consider the following three models for the random fluctuations of the position and orientation of the UAV: \textit{i) Independent Gaussian Fluctuations:} Since the fluctuations are in general the result of many contributing factors, such as random air fluctuations in the atmosphere around the UAV and internal vibrations of the UAV, by invoking the central limit theorem, we model the resulting
fluctuations of the UAV as Gaussian distributed random variables (RVs) \cite{Steve_pointing_error,Alouini_Pointing}.  \textit{ii) Correlated Gaussian Fluctuations:} This more general model allows for correlations that can be the result of e.g. wind causing the UAV to have stronger fluctuations in a certain direction. \textit{iii) Correlated Uniform Fluctuations:} We also consider uniformly distributed fluctuations which may better model the characteristics of UAV fluctuations with large variance  than Gaussian distributed fluctuations  \cite{Uniform1,Uniform2,Uniform3}. Independent Gaussian, correlated Gaussian, and correlated uniform  fluctuations are expected to be suitable models for calm, weakly windy, and strongly windy weather conditions, respectively.
\item We derive novel statistical models for the GML for each fluctuation scenario. Moreover, we simplify the derived closed-form expressions for some special cases, e.g., when the beam is orthogonal to the receiver lens plane, to obtain further insight.  
\item Based on the developed statistical GML models, we analyze the performance of a UAV-based FSO link in terms of outage probability and ergodic rate. In particular, we assume that the impact of the GML is dominant compared to atmospheric turbulence induced fading. This is a valid assumption when the distance between the UAV and the CU is on the order of several hundred meters, as is validated by simulations in Section~VI, cf. Figs.~\ref{Fig:Outage} and \ref{Fig:Rate}. Next, we derive analytical expressions for the outage probability and ergodic rate of the considered system and analyze their asymptotic behavior for high signal-to-noise ratios (SNRs) for the three statistical GML models. 
\item Simulations are used to validate our derivations and show the impact of the system parameters, e.g., the non-orthogonality of the optical beam, the variance of the fluctuations, and the beam width, on system performance. Our results reveal the existence of a trade-off between the outage probability and the amount of the average (and the maximum) power collected by the receiver. More specifically, when the variance of the fluctuations is large, a wider beam is preferable to avoid outages although this decreases the average (and the maximum) power collected at the receiver. On the other hand, when the variance of the fluctuations is small, a narrower beam is preferable since this increases the amount of power collected by the receiver lens, see  Fig.~\ref{Fig:CDF_Unif}.
\end{itemize}

\begin{figure*}[t]
	\centering
	\includegraphics[width=1\linewidth]{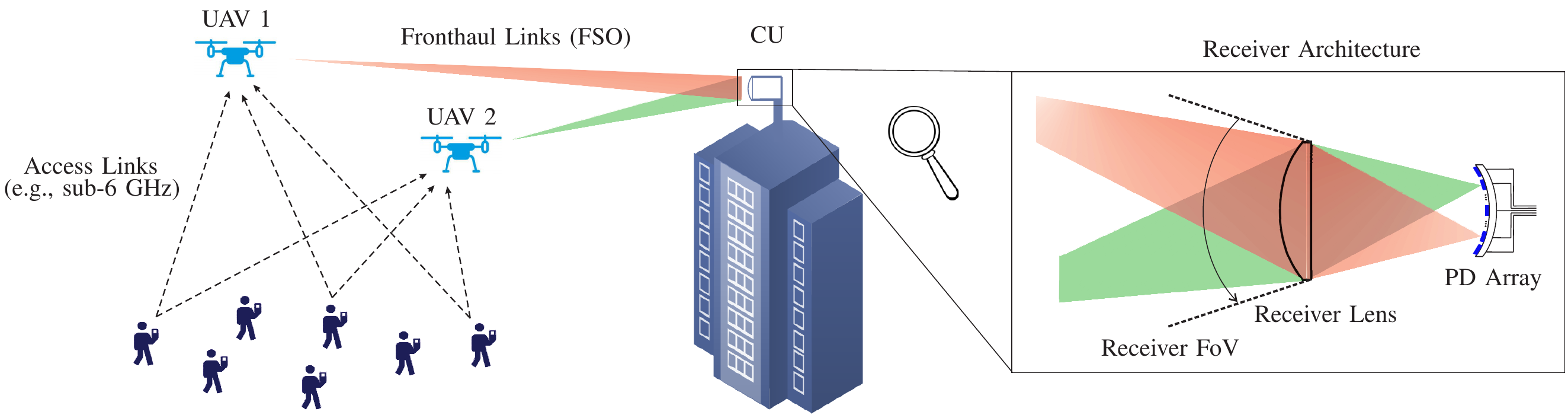}
	\caption{Proposed UAV-based communication system where the UAVs communicate with the mobile users via an RF multiple-access link and with the CU via FSO fronthaul links. The CU is equipped with a lens placed in front of an array of PDs. This architecture yields a large FoV and enables the CU to separate signals coming from different spatial angles (i.e., from one moving UAV at different positions or multiple UAVs at different locations).}
	\label{Fig:LensArray}
\end{figure*}

The remainder of this paper is organized as follows: The system and channel models are presented in Section II. In Section III, we develop the conditional GML model, and in Section IV, we derive statistical GML models for three different fluctuation scenarios. In Section~V, we analyze the performance of a UAV-based FSO link using the developed channel models. In Section VI, we present simulation results, and Section VII concludes the paper.

\textit{ Notations:} Boldface lower-case and upper-case letters are reserved for vectors and matrices, respectively. $\mathbbmss{E}\{\cdot\}$, $(\cdot)^{\mathsf{T}}$, and $\|\cdot\|$ denote expectation, the transpose of a matrix,  and the $l_2$-norm of a vector, respectively.  $\mathbb{R}$ and $\mathbb{R}^+$ denote the sets of  real and positive real numbers, respectively. $\mathbf{I}$ represents the identity matrix  and $\mathrm{diag}\{a_1,\dots,a_n\}$ denotes a diagonal matrix with $a_1,\dots,a_n$ on its main diagonal.  $\ln(\cdot)$, $\mathrm{erf}(\cdot)$, $Q(\cdot)$, and $Q(\cdot,\cdot)$ denote the natural logarithm, the error function, the Gaussian Q-function, and the first-order Marcum Q-function, respectively. $\mathbf{a}\sim\mathcal{N}(\boldsymbol{\mu},\boldsymbol{\Sigma})$ is used to indicate that~$\mathbf{a}$ is a multivariate Gaussian random vector with mean vector $\boldsymbol{\mu}$ and covariance matrix~$\boldsymbol{\Sigma}$ and $b\sim\mathcal{U}(a,b)$ means that RV $b$ is uniformly distributed in interval $[a,b]$. Finally, $\mathbf{a}\cdot\mathbf{b}$ and $\mathbf{a}\times\mathbf{b}$ denote the dot and cross products of vectors $\mathbf{a}$ and $\mathbf{b}$, respectively.
 
\section{System and Channel Models}

\subsection{System Model} 
We consider a UAV-based uplink transmission  where the UAV communicates with mobile users via an RF multiple-access link (e.g., using sub-6 GHz bands) and with the CU via an FSO fronthaul link, see Fig.~\ref{Fig:LensArray}. The focus of this paper is on the fronthaul communication between the UAV and the CU. In particular, we assume that the UAV is equipped with an aperture FSO transmitter pointing towards the CU, which detects the received optical power. To avoid the requirement of a mechanical adjustment of the orientation of the receiver at the CU, we assume that the receiver is equipped with a lens placed in front of an array of PDs \cite{droz2014photodetector,hahn2010fiber,ghassemlooy2019optical}, cf. Fig.~\ref{Fig:LensArray}. The lens separates the signals coming from different spatial angles and focuses them on corresponding PDs, respectively. The large overall FoV of the PD array enables the CU to receive data when a moving UAV is at different positions. Moreover, this architecture allows the CU to simultaneously receive data from multiple UAVs which are spatially separated but are still within the receiver FoV \cite{hahn2010fiber}. The overall receiver FoV and the FoV of the individual PDs are design parameters which depend on the specific implementation \cite{droz2014photodetector,hahn2010fiber}. As mentioned before, the main goal of this paper is to develop a mathematical framework that models the impact of the fluctuations of the position and orientation of a hovering UAV on the FSO channel quality.  Therefore, we assume an ideal PD array where one of the PDs is able to collect the entire optical power flux into the receiver lens (albeit with a fixed efficiency/responsivity factor, see Section~\ref{Sec:FSOchannel}).


\begin{figure*}[t]
\centering
\includegraphics[width=0.6\linewidth]{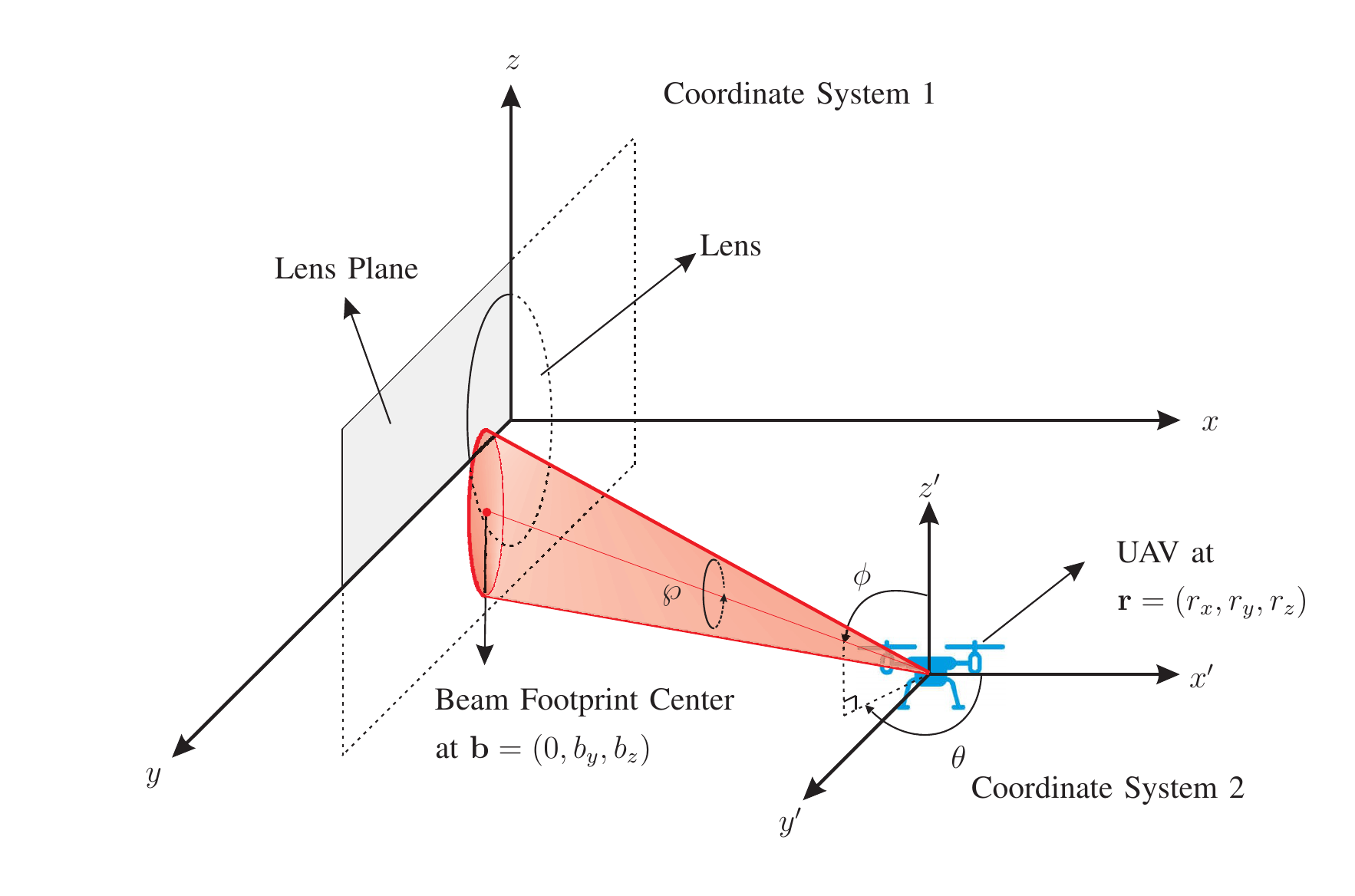}
\caption{CU and UAV coordinate systems.}
\label{Fig:Nodes}
\end{figure*}
To characterize an object in three dimensions, \textit{at most} six independent variables are needed: three variables to specify the position of a reference point of the object and another three to quantify its orientation. Next, we define the position and orientation of the UAV and the CU.

\subsubsection{CU} The CU is a fixed node located at the top of a building\footnote{The CU may not be stable due to building sway \cite{Steve_pointing_error,Alouini_Pointing}. Nevertheless, since only the relative movement of UAV and CU affects the FSO channel quality, we assume that the CU is fixed and only the UAV moves. Note that, in practice, the movement of the CU is negligible compared to the movement of the UAV.}. Without loss of generality, we choose the center of the receiver lens as the reference point, which is located in the origin of the Cartesian coordinate system $(x,y,z)=(0,0,0)$. This coordinate system is referred to as Coordinate System~1, cf. Fig.~\ref{Fig:Nodes}. Moreover, we assume a circular lens of radius $r_0$. Note that it suffices to characterize the plane where the receiver lens lies in order to specify its orientation. Here, without loss of generality, we assume that the lens lies in the $y-z$ plane at $x=0$.

\subsubsection{UAV} For the communication system under consideration, the parameters that directly affect the FSO channel are the position of the laser source of the UAV and the direction of the laser beam. Therefore, without loss of generality, we refer to them as the position and orientation of the UAV, respectively. Furthermore, we assume that the UAV is in the hovering state. However, in practice, the position and orientation of the UAV are not perfectly constant in the hovering state \cite{Wind1,Wind2,Wind3} and thus, they are modeled as RVs. In particular, let $\mathbf{r}=(r_x,r_y,r_z)$ and $\boldsymbol{\omega} = (\theta,\phi,\wp)$ denote the vectors containing the random position and orientation variables of the UAV, respectively. Without loss of generality, in order to simplify the analysis, we define vector $\mathbf{r}$ w.r.t. Coordinate System~1, whereas we use the following coordinate system for $\boldsymbol{\omega}$: For a given vector $\mathbf{r}$, we define Coordinate System~2 with $\mathbf{r}$ as its origin and axes $x'$, $y'$, and $z'$ that are parallel to the $x$, $y$, and $z$ axes of Coordinate System~1, respectively, cf. Fig.~\ref{Fig:Nodes}. We use variables $\theta$ and $\phi$ to determine the direction of the laser beam in a spherical representation of Coordinate System~2. In particular, $\theta\in[0,2\pi]$ denotes the angle between the projection of the beam vector onto the $x'-y'$ plane and the $x'$ axis; and $\phi\in[0,\pi]$ represents the angle between the beam vector and the $z'$ axis. The third orientation variable $\wp$ is used to quantify the rotation around the beam vector. This representation of the orientation variables has two advantages. First, variable  $\boldsymbol{\omega}$ does not change if position $\mathbf{r}$ changes, i.e., the position and orientation variables are independent. Second, a rotation around the beam line does not affect the signal at the PD assuming rotational beam symmetry. Therefore, the value of $\wp$ is irrelevant for the analysis, and hereafter, for simplicity, we drop $\wp$ and use $\boldsymbol{\omega} = (\theta,\phi)$ as the random vector of the orientation variable.

\subsection{FSO Channel Model}\label{Sec:FSOchannel}
We assume an intensity modulation/direct detection (IM/DD) FSO system, where the PD responds to changes in the received optical signal  power  \cite{FSO_Survey_Murat}. Moreover, we assume that  noise caused by background illumination is the dominant noise source at the PD \cite{FSO_Survey_Murat,Kahn_SigIndepNoise}. Hence, in our model, the noise is independent from the signal.
 The received signal at the CU is given by
 \begin{IEEEeqnarray}{lll}\label{Eq:signal}
 y_s=hx_s+n,
 \end{IEEEeqnarray}
 where $x_s\in \mathbb{R}^+$ is the transmitted optical symbol (intensity), $n\in \mathbb{R}$ is the zero-mean real-valued additive white Gaussian shot noise with variance $\sigma_n^2$ caused by background illumination at the CU, and $h\in \mathbb{R}^+$ denotes the FSO channel coefficient. Moreover, we assume an average power constraint $\mathbbmss{E}\{x_s\}\leq{P}$. The FSO channel coefficient, $h$, is affected by several phenomena and can be modeled as \cite{Steve_pointing_error} 
\begin{IEEEeqnarray}{lll}\label{Eq:channel}
h=\eta h_p h_a h_g,
\end{IEEEeqnarray}
where $\eta$ is the responsivity of the PD and $h_p$, $h_a$, and $h_g$ represent the atmospheric loss, atmospheric turbulence induced fading, and GML, respectively. In particular, the atmospheric loss, $h_p$, is deterministic and represents the power loss over a propagation path due to absorption and scattering of the light by particles in the atmosphere. It is modeled as \cite{Schober,FSO_Vahid}
\begin{IEEEeqnarray}{rll}\label{Eq:Pathloss}
h_p=10^{-\kappa L/10},
\end{IEEEeqnarray}
where $L$ is the distance between the UAV and the CU and $\kappa$ [m$^{-1}$] denotes the weather-dependent attenuation constant of the FSO link. For clear air, haze, light fog, moderate fog, and heavy fog, the typical values of $\kappa$ are $\{0.43,4.2,20,42.2,125\}\times10^{-3}$~m$^{-1}$, respectively  \cite{FSO_Vahid}.

The atmospheric turbulence, $h_a$, is an RV and induced 
by inhomogeneities in the temperature and the pressure of the atmosphere. It is typically modeled as log-normal (LN) and Gamma-Gamma (GG) 
distributed RV for weak and moderate-to-strong turbulence conditions \cite{Steve_pointing_error}, respectively.
For the considered system, the distance between the UAV and the CU is typically on the order of several hundred meters. In this regime, the atmospheric turbulence is moderate and its impact is negligible compared to that of the GML. To verify this claim, let us consider the pessimistic GG fading model, i.e., $h_a\sim\mathcal{GG}(\alpha,\beta)$, with fading parameters $\alpha$ and $\beta$ \cite{GG_Model}. In particular, for GG fading, $h_a$ is modeled as the product of two independent Gamma random variables $h_a^{(1)}\sim\mathcal{G}(\alpha,\alpha)$ and $h_a^{(2)}\sim\mathcal{G}(\beta,\beta)$, which represent irradiance fluctuations	arising from large- and small-scale turbulences, respectively \cite{GG_Model}. Parameters $\alpha$ and $\beta$ are the inverse of the variances of  $h_a^{(1)}$ and  $h_a^{(2)}$, respectively, and are given by \cite{Alouini_Pointing}
\iftoggle{OneColumn}{%
\begin{IEEEeqnarray}{rll}\label{Eq:alphabeta}
	 \alpha=\left[\exp\bigg(\tfrac{0.49\sigma_R^2}{(1+1.11\sigma_R^{12/5})^{7/6}}\bigg)-1\right]^{-1}\text{and}\,\,\, 
	\beta=\left[\exp\bigg(\tfrac{0.51\sigma_R^2}{(1+0.69\sigma_R^{12/5})^{5/6}}\bigg)-1\right]^{-1}.\quad
\end{IEEEeqnarray}
}{%
\begin{IEEEeqnarray}{rll}\label{Eq:alphabeta}
	&\alpha=\left[\exp\left(\tfrac{0.49\sigma_R^2}{\left(1+1.11\sigma_R^{12/5}\right)^{7/6}}\right)-1\right]^{-1},\nonumber\\
	&\beta=\left[\exp\left(\tfrac{0.51\sigma_R^2}{\left(1+0.69\sigma_R^{12/5}\right)^{5/6}}\right)-1\right]^{-1}.\quad
\end{IEEEeqnarray}
}
In \eqref{Eq:alphabeta}, $\sigma_R^2=1.23C_n^2k^{7/6}L^{11/6}$ is the Rytov variance, $k=2\pi/\lambda$, where $\lambda$ [m] denotes the optical wavelength, and $C_n^2\approx C_0^2\exp\left(-\frac{h_d}{100}\right)$ [m$^{-\frac{2}{3}}$] is the index of refraction structure parameter, where $h_d$ is the operating height of the UAV and $C_0^2=1.7\times10^{-14}$~m$^{-\frac{2}{3}}$ is the nominal value of the refractive index at the ground \cite{Alouini_Pointing}. For typical system parameters, the variance of $h_a$, i.e., $\frac{1}{\alpha}+\frac{1}{\beta}+\frac{1}{\alpha\beta}$, is very small  (e.g., $3\times 10^{-2}$ for $L=500$~m, $h_d=120$~m, and $\lambda=1550$~nm). Therefore, we approximate $h_a$ by its mean value, i.e., $h_a\approx \mathbbmss{E}\{h_a\}=\mathbbmss{E}\{h_a^{(1)}\}\mathbbmss{E}\{h_a^{(2)}\}=\frac{\alpha}{\alpha}\times\frac{\beta}{\beta}=1$. We verify this assumption by simulations in Section~VI, cf. Figs.~\ref{Fig:Outage} and \ref{Fig:Rate}.


The GML, $h_g$, is caused by the divergence of the optical beam between the transmitter and the receiver lens and the misalignment of the laser beam line and the center of the lens \cite{FSO_Survey_Murat,FSO_Vahid}. Fluctuations of the position and orientation of the UAV lead to a random GML, $h_g$. In the following, we first derive a conditional model for the GML and then, we develop statistical models for independent Gaussian, correlated Gaussian, and uniformly distributed fluctuations. 



\section{The Conditional GML Model}\label{Sec:hg_det}
In this section, we derive the channel parameter $h_g$ for a given state of the UAV, i.e., for given $\mathbf{r}$~and~$\boldsymbol{\omega}$. 

\subsection{Center of the Beam Footprint}

The line of the beam can be represented in Cartesian Coordinate System~1 as 
\begin{IEEEeqnarray}{lll} \label{Eq:BeamLine}
(x,y,z) = \mathbf{r} + \jmath \mathbf{d},
\end{IEEEeqnarray}
where $\jmath $ is an arbitrary real number and $\mathbf{d}=(d_x,d_y,d_z)$ denotes the beam direction, which can be written as a function of $\theta$ and $\phi$ as 
\begin{IEEEeqnarray}{lll} \label{Eq:d_angle}
\mathbf{d}=\big(\sin\phi\cos\theta,\sin\phi\sin\theta,\cos\phi\big).
\end{IEEEeqnarray}
The center of the beam footprint on the receiver lens can be obtained as the intersection point of the line of the laser beam and the lens plane, $x=0$. Denoting the center of the footprint of the beam on the lens as $\mathbf{b}=(b_x,b_y,b_z)$, then
\begin{IEEEeqnarray}{lll} \label{Eq:FootPrint_Center}
\mathbf{b} = \Big(0,r_y-r_x\tan\theta,r_z-r_x\frac{\cot\phi}{\cos\theta}\Big).
\end{IEEEeqnarray}

\subsection{Power Density on the Lens Plane}

We assume a Gaussian beam, which dictates that the power density distribution across any plane perpendicular to the direction of the wave propagation follows a Gaussian profile \cite{FSO_Survey_Murat,Steve_pointing_error}. In particular, we consider a perpendicular plane where the distance between the center of the beam footprint on the plane and the laser source is denoted by $L$. Then, the power density for any point on this perpendicular plane with distance $l$ from the center of the beam footprint is given by \cite{Steve_pointing_error}
\begin{IEEEeqnarray}{lll} \label{Eq:PowerOrthogonal}
I^{\mathrm{orth}}(l;L) = \frac{2}{\pi w_L^2}\exp\left(-\frac{2l^2}{w_L^2}\right),
\end{IEEEeqnarray}
where $w_L$ [m] is the beam width at distance $L$ and can be evaluated as
\begin{IEEEeqnarray}{lll} \label{Eq:BeamWidth}
w_L = w_0\sqrt{ 1+\left(1+\frac{2w_0^2}{\rho^2(L)}\right)\left(\frac{\lambda L}{\pi w_0^2}\right)^2}.
\end{IEEEeqnarray}
 For the case where the beam propagates in the $x$ direction, $l=\sqrt{\tilde{y}^2+\tilde{z}^2}$ holds where $\tilde{y}=y-b_y$ and $\tilde{z}=z-b_z$. In (\ref{Eq:BeamWidth}), $w_0$ [m] denotes the beam waist radius and $\rho(L)=(0.55C_n^2k^2L)^{-3/5}$ [m] is referred to as the coherence length.
Recall that for the problem at hand, the plane of the receiver lens is not necessarily orthogonal to the beam direction. For this case, the power density on the lens plane, denoted by $I(y,z)$, is given in the following lemma.

\begin{lem}\label{Lem:PowerDensity}
Under the mild conditions $\Vert \mathbf{r} \Vert \gg \Vert \mathbf{b} \Vert$ and $\Vert \mathbf{r} \Vert \gg   \Vert (y,z) \Vert$, the power density at point $(y,z)$ on the PD plane is given by
\iftoggle{OneColumn}{%
\begin{IEEEeqnarray}{lll} \label{Eq:Intensity}
I(y,z)&=\sin\psi I^{\mathrm{orth}}\big(l(\boldsymbol{\omega},y,z);L(\mathbf{r})\big) 
&=\frac{2\sin\psi}{\pi w^2_L}\exp\Big(\frac{-2}{ w^2_L}(\rho_{y}\tilde{y}^2+\rho_{z}\tilde{z}^2
+ 2\rho_{yz}\tilde{y}\tilde{z})\Big),\quad\,\,\,\,
\end{IEEEeqnarray}
}{%
\begin{IEEEeqnarray}{lll} \label{Eq:Intensity}
I(y,z)&=\sin\psi I^{\mathrm{orth}}\big(l(\boldsymbol{\omega},y,z);L(\mathbf{r})\big) \nonumber\\
&=\frac{2\sin\psi}{\pi w^2_L}
\exp\Big(\frac{-2}{ w^2_L}(\rho_{y}\tilde{y}^2+\rho_{z}\tilde{z}^2
+ 2\rho_{yz}\tilde{y}\tilde{z})\Big),\quad\,\,\,\,
\end{IEEEeqnarray}
}
where $\psi=\sin^{-1}(\sin\phi \cos\theta)$ is the angle between the beam line and the lens plane, $l(\boldsymbol{\omega},y,z)=\sqrt{\rho_{y}\tilde{y}^2+\rho_{z}\tilde{z}^2
	+ 2\rho_{yz}\tilde{y}\tilde{z}}$, $L(\mathbf{r})=\Vert \mathbf{r} \Vert$, and $I^{\mathrm{orth}}(\cdot;\cdot)$ is given by (\ref{Eq:PowerOrthogonal}). Moreover, $\rho_{y} = \cos^2\phi+\sin^2\phi\cos^2\theta$,
$\rho_{z} = \sin^2\phi$, and
$\rho_{yz} = -\cos\phi\sin\phi\sin\theta$.
\end{lem}
\begin{IEEEproof}
The proof is given in Appendix~\ref{App:Lem_PowerDensity}.
\end{IEEEproof}

Note that the conditions under which (\ref{Eq:Intensity}) holds are met in practice since, for typical FSO links, $\Vert \mathbf{r} \Vert$ is on the order of several hundred meters, whereas $\Vert \mathbf{b} \Vert$ and $\Vert (y,z) \Vert$  are on the order of a few centimeters.

\subsection{GML}

The fraction of power collected by the receiver lens, denoted by $h_g(\mathbf{r},\boldsymbol{\omega})$, can be obtained by integrating the power density derived in Lemma~\ref{Lem:PowerDensity} over the lens area. This leads to
\iftoggle{OneColumn}{%
\begin{IEEEeqnarray}{lll}\label{Eq:PowerPhotoDetector} 
h_g(\mathbf{r},\boldsymbol{\omega}) =  \underset{(y,z)\in\mathcal{A}}{\iint} 
I(y,z) \mathrm{d}y\mathrm{d}z, 
\end{IEEEeqnarray}
}{%
\begin{IEEEeqnarray}{lll}\label{Eq:PowerPhotoDetector} 
h_g(\mathbf{r},\boldsymbol{\omega}) =  \underset{(y,z)\in\mathcal{A}}{\iint} 
I(y,z) \mathrm{d}y\mathrm{d}z,
\end{IEEEeqnarray}
}
where $I(y,z) $ is given in (\ref{Eq:Intensity}) and $\mathcal{A}=\left\{(y,z)|y^2+z^2\leq r_0^2\right\}$ is the set of $(y,z)$ within the lens area. The exact value of the integral in (\ref{Eq:PowerPhotoDetector}) cannot be obtained in closed form. Instead, in the following theorem, we provide an upper and a lower bound on $h_g(\mathbf{r},\boldsymbol{\omega})$. 

\begin{theo}\label{Theo:Power}
Using Lemma~\ref{Lem:PowerDensity}, $h_g(\mathbf{r},\boldsymbol{\omega})$ can be lower and upper bounded by 
\iftoggle{OneColumn}{%
\begin{IEEEeqnarray}{lll} \label{Eq:upper_lower_bound}
 h_g^{\mathrm{low}}(\mathbf{r},\boldsymbol{\omega}) =\frac{2\sin \psi }{\pi w^2_L}
 \underset{(y,z)\in\mathcal{A}}{\iint} 
\exp\left(-\frac{2}{ w^2_L}\Big((y-u)^2+\sin^2\phi\cos^2\theta{z}^2\Big)\right) 
\mathrm{d}y\mathrm{d}z\quad\text{and}\IEEEyesnumber\IEEEyessubnumber\qquad\\
 h_g^{\mathrm{upp}}(\mathbf{r},\boldsymbol{\omega}) =\frac{2\sin \psi }{\pi w^2_L}
 \underset{(y,z)\in\mathcal{A}}{\iint} 
\exp\left(-\frac{2}{ w^2_L}\Big(\sin^2\phi\cos^2\theta(y-u)^2+{z}^2\Big)\right)
\mathrm{d}y\mathrm{d}z, \,\, \IEEEyessubnumber\quad
\end{IEEEeqnarray}
}{%
\begin{IEEEeqnarray}{lll} \label{Eq:upper_lower_bound}
 h_g^{\mathrm{low}}(\mathbf{r},\boldsymbol{\omega}) =\frac{2\sin \psi }{\pi w^2_L}\underset{(y,z)\in\mathcal{A}}{\iint} \nonumber\\
\exp\bigg(-\frac{2}{ w^2_L}\Big((y-u)^2+\sin^2\phi\cos^2\theta{z}^2\Big)\bigg) 
\mathrm{d}y\mathrm{d}z\IEEEyesnumber\IEEEyessubnumber\\
\text{and} \nonumber\\
 h_g^{\mathrm{upp}}(\mathbf{r},\boldsymbol{\omega}) =\frac{2\sin \psi }{\pi w^2_L}\underset{(y,z)\in\mathcal{A}}{\iint} \nonumber\\
\exp\bigg(-\frac{2}{ w^2_L}\Big(\sin^2\phi\cos^2\theta(y-u)^2+{z}^2\Big)\bigg)
\mathrm{d}y\mathrm{d}z, \,\, \IEEEyessubnumber\quad
\end{IEEEeqnarray}
}
respectively. Here, $u=\|\mathbf{b}\|$ denotes the distance between the origin and the center of the beam footprint, i.e., the misalignment.  
\end{theo}
\begin{IEEEproof}
The proof is given in Appendix~\ref{App:Integral}.
\end{IEEEproof}

\begin{figure}[t]
	\centering
	\includegraphics[width=1\linewidth]{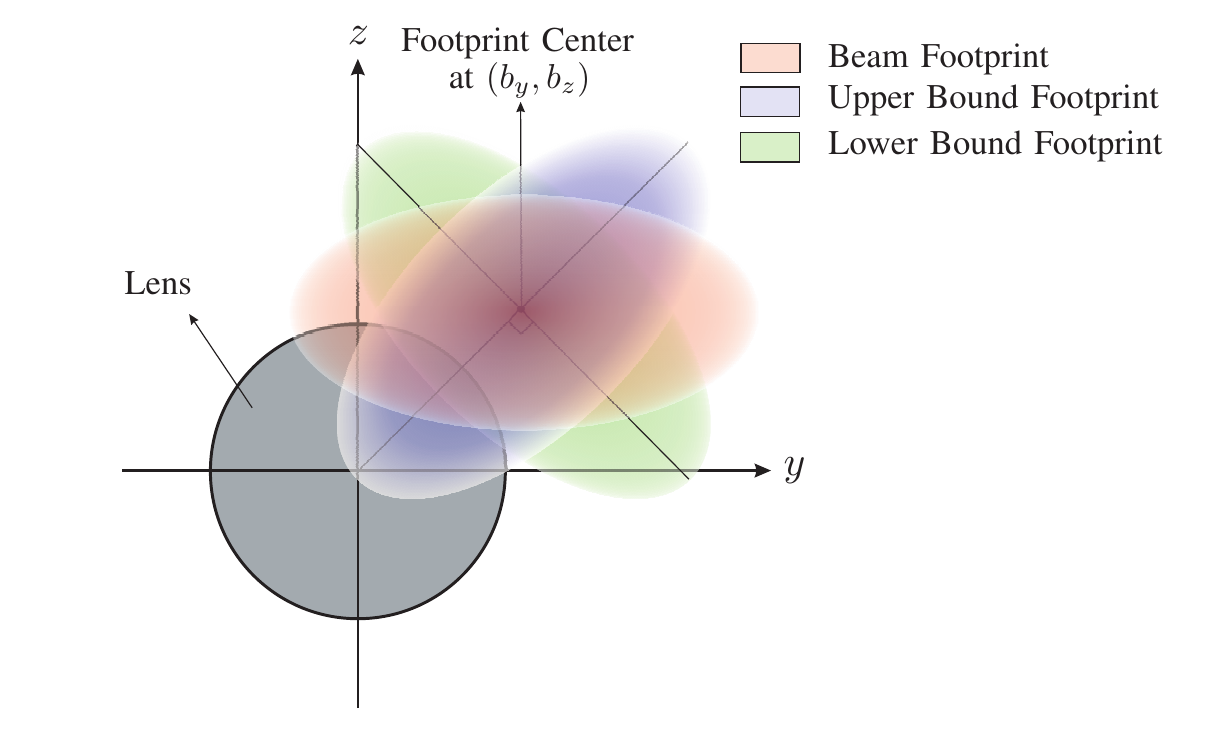}
	\caption{Beam footprint on the receiver lens plane and the footprints that are used to derive the upper and lower bounds for the GML.}
	\label{Fig:Contour}
\end{figure}

\begin{remk}
We use Fig.~\ref{Fig:Contour} to illustrate the basic idea behind the upper and lower bounds proposed in Theorem~\ref{Theo:Power}. In particular, unlike the case where the optical beam is orthogonal to the lens plane and the power density contours are circles \cite{Steve_pointing_error}, when the optical beam is non-orthogonal to the lens plane, the power density contours are \textit{rotated ellipses}, e.g., the red ellipse in Fig.~\ref{Fig:Contour}. We have derived the lower bound assuming a footprint that is a rotated ellipse, whose major axis is perpendicular to the line connecting the center of the footprint and the origin, i.e., the green ellipse in Fig.~\ref{Fig:Contour}. Moreover, for the upper bound, the footprint is a rotated ellipse, whose minor axis is perpendicular to the line connecting the center of the footprint and the origin, i.e., the purple ellipse in Fig.~\ref{Fig:Contour}. In the special case where the major (minor) axis of the \textit{original} power density contour is perpendicular to the line connecting the center of the footprint and the origin, the lower (upper) bound is identical to the exact GML.
\end{remk}

The integrals in \eqref{Eq:upper_lower_bound} cannot be evaluated in closed-form. Even for the case where the beam line is orthogonal to the lens plane (as is the case for conventional FSO systems \cite{Steve_pointing_error}), the exact value of $h_g(\mathbf{r},\boldsymbol{\omega})$ is cumbersome and provides little insight. Therefore, in \cite{Steve_pointing_error}, the authors proposed an approximation for conventional FSO systems, which was shown to be very accurate for $\frac{w_L}{r_0}\geq 6$ and has been widely used by other authors subsequently \cite{Steve_MISO_FSO,George_Pointing_error,Alouini_Pointing,
Khalighi_FSO_UAV_ChannelModel}. The proposed bounds in Theorem~\ref{Theo:Power} have two main advantages. First, for the special case where the beam line is orthogonal to the lens plane, the upper and lower bounds coincide with the exact $h_g(\mathbf{r},\boldsymbol{\omega})$. Second, the form of the integrals in (\ref{Eq:upper_lower_bound}) allows to employ the same technique as in \cite[Appendix]{Steve_pointing_error} in order to obtain accurate approximations. In particular, as shown in detail in Appendix~\ref{App:Low_Upp_approx}, we approximate $h_g^{\mathrm{low}}(\mathbf{r},\boldsymbol{\omega})$ and $h_g^{\mathrm{upp}}(\mathbf{r},\boldsymbol{\omega})$ in (\ref{Eq:upper_lower_bound}) with $\widetilde{h}_g^{\mathrm{low}}(\mathbf{r},\boldsymbol{\omega})$ and $\widetilde{h}_g^{\mathrm{upp}}(\mathbf{r},\boldsymbol{\omega})$, respectively, as follows
\iftoggle{OneColumn}{%
\begin{IEEEeqnarray}{lll} \label{Eq:upper_lower_cal}
\widetilde{h}_g^{\mathrm{low}}(\mathbf{r},\boldsymbol{\omega}) = A_0 \exp\left(-\frac{2u^2}{t_1w^2_L}\right),\quad
 \widetilde{h}_g^{\mathrm{upp}}(\mathbf{r},\boldsymbol{\omega}) = A_0 \exp\left(-\frac{2u^2}{t_2w^2_L}\right),
\end{IEEEeqnarray}
}{%
\begin{IEEEeqnarray}{lll} \label{Eq:upper_lower_cal}
\widetilde{h}_g^{\mathrm{low}}(\mathbf{r},\boldsymbol{\omega}) = A_0 \exp\left(-\frac{2u^2}{t_1w^2_L}\right)\IEEEyesnumber\IEEEyessubnumber\\
 \widetilde{h}_g^{\mathrm{upp}}(\mathbf{r},\boldsymbol{\omega}) = A_0 \exp\left(-\frac{2u^2}{t_2w^2_L}\right),\IEEEyessubnumber
\end{IEEEeqnarray}
}
 where $t_1=\frac{\sqrt{\pi}\mathrm{erf}(\nu_1)}{2\nu_1\exp(-\nu_1^2)}$, $\nu_{1}=\frac{r_0}{w_L}\sqrt{\frac{\pi}{2}}$, $t_2=\frac{\sqrt{\pi}\mathrm{erf}(\nu_2)}{2\nu_2\exp(-\nu_2^2)\sin^2\phi\cos^2\theta}$, and  $\nu_{2}=\nu_{1}|\sin\phi\cos\theta|$. Moreover, $A_0$ denotes the maximum fraction of optical power captured by the receiver lens at $u=0$ and is given by
\begin{IEEEeqnarray}{lll}\label{Eq:A_0}
A_0=\mathrm{erf}(\nu_{1})\mathrm{erf}(\nu_{2}).
\end{IEEEeqnarray}
Note that $A_0$ is inversely proportional to $\frac{w_L}{r_0}$, which means that, as expected, the wider the beam footprint w.r.t. the lens is, the smaller the amount of power that can be collected by the lens. 
 The only difference between the approximated lower and upper bounds in (\ref{Eq:upper_lower_cal}) are the factors $t_1$ and $t_2$. This motivates us to propose the following  approximation for the GML
\begin{IEEEeqnarray}{lll} \label{Eq:hg_mean}
h_g(\mathbf{r},\boldsymbol{\omega}) \approx A_0 \exp\left(-\frac{2u^2}{tw^2_L}\right),
\end{IEEEeqnarray}
where $t\in[t_1,t_2]$. In (\ref{Eq:hg_mean}), $h_g(\mathbf{r},\boldsymbol{\omega})$ comprises two parts, namely $A_0$, which affects the geometric loss and $\exp\left(-\frac{2u^2}{tw^2_L}\right)$, which represents the misalignment attenuation when $u\neq 0$. 

In the following, instead of considering the approximate upper and lower bounds in (\ref{Eq:upper_lower_cal}), we employ the general approximation in (\ref{Eq:hg_mean}) for  statistical analysis. One can choose $t$ in (\ref{Eq:hg_mean}) equal to $t_1$ and $t_2$ to obtain the lower and upper bounds, respectively. Alternatively, $t$ can be chosen as the arithmetic mean $\frac{t_1+t_2}{2}$ or the geometric mean $\sqrt{t_1t_2}$ to compromise between the lower and upper bounds. 
Our results in Section \ref{Sec:Sim} show that the approximation in (\ref{Eq:hg_mean}) yields an accurate approximation of $h_g(\mathbf{r},\boldsymbol{\omega})$ for both $t=\frac{t_1+t_2}{2}$ and $t=\sqrt{t_1t_2}$ for the practical range of system parameters, cf. Fig.~\ref{Fig:Bound}.

\section{Statistical Models for the GML}

In \eqref{Eq:hg_mean}, we provided an approximate closed-form expression for the GML $h_g(\mathbf{r},\boldsymbol{\omega})$ for given values of $\mathbf{r}$ and $\boldsymbol{\omega}$. However, in practice, the position and orientation of a hovering UAV fluctuate randomly, and hence, $\mathbf{r}$ and $\boldsymbol{\omega}$ are RVs. In the following, we first present three fluctuation scenarios for RVs $\mathbf{r}$ and $\boldsymbol{\omega}$ and then derive the corresponding statistical GML models.

\subsection{Models for the Random Position and Orientation Fluctuations of the UAV}

As mentioned above, the position and orientation of the UAV randomly fluctuate over time. In other words, a hovering UAV is not perfectly stable \cite{UAV_FSO}. Therefore, an active control mechanism is needed to persistently keep the laser beam and the receiver lens aligned (see \cite{Controller_FSO_Komaee,Controller_FSO_Arnon,
Controller_FSO_Yuksel,Robust_laser_tracking}). For ideal tracking, the center of the beam footprint coincides with the center of the receiver lens, i.e., $u=0$. Nevertheless, in practical systems, misalignment due to tracking errors exists for several reasons. For instance, the control system requires some time to compensate for a misalignment or this system is not perfectly accurate, while compensating for a misalignment. Moreover, in UAVs, there is an error associated with wind estimation, i.e., the power and direction of wind\cite{Wind2}, and therefore, the impact of wind cannot be fully compensated. In fact, tracking errors exist even in conventional FSO systems, where the transceivers are mounted on top of buildings and misalignment originates from building sway. However, for UAV-based FSO links, such tracking errors are expected to be more severe, due to the inherent instability of hovering UAVs. Therefore, for the development of a channel model for UAV-based FSO links, statistical models for the position and orientation of the UAV are needed.  

Let us define vectors $\boldsymbol{\mu}_{\mathbf{r}}=(\mu_x,\mu_y,\mu_z)$ and $\boldsymbol{\mu}_{\boldsymbol{\omega}}=(\mu_{\theta},\mu_{\phi})$, which denote the means of random vectors $\mathbf{r}$ and $\boldsymbol{\omega}$, respectively. Furthermore, we define the zero-mean random vectors $\boldsymbol{\epsilon}_{\mathbf{r}}=({\epsilon}_{x},{\epsilon}_{y},{\epsilon}_{z})$ and $\boldsymbol{\epsilon}_{\boldsymbol{\omega}}=({\epsilon}_{\theta},{\epsilon}_{\phi})$ to model the fluctuations of the position and orientation of the UAV, respectively. Therefore, the position and orientation of the UAV are respectively given by 
\begin{IEEEeqnarray}{lll} \label{Eq:r_omega_RV}
\mathbf{r} = \boldsymbol{\mu}_{\mathbf{r}}+\boldsymbol{\epsilon}_{\mathbf{r}}
 \quad \text{and} \quad
\boldsymbol{\omega}= \boldsymbol{\mu}_{\boldsymbol{\omega}}+\boldsymbol{\epsilon}_{\boldsymbol{\omega}}.
\end{IEEEeqnarray}
 Since the GML is a function of $\boldsymbol{\epsilon}_{\mathbf{r}}$ and  $\boldsymbol{\epsilon}_{\boldsymbol{\omega}}$, the distribution of $\boldsymbol{\epsilon}_{\mathbf{r}}$ and  $\boldsymbol{\epsilon}_{\boldsymbol{\omega}}$ determines the distribution of the GML. Hence, adopting appropriate distributions for fluctuations $\boldsymbol{\epsilon}_{\mathbf{r}}$ and  $\boldsymbol{\epsilon}_{\boldsymbol{\omega}}$ is important for developing a realistic statistical model for the GML. Hereby, independent Gaussian, correlated Gaussian, and correlated uniformly distributed fluctuations may serve as models for calm, weakly windy, and strongly windy weather conditions, respectively. In the following, we discuss the first and second moments of RVs $\mathbf{r}$ and $\boldsymbol{\omega}$. 

\subsubsection{First Moments of RVs}

Since the UAV is supposed to hover above the area where the users are located,  $\boldsymbol{\mu}_{\mathbf{r}}$  depends on the location of the users as well as the desired operating height of the UAV. Given $\boldsymbol{\mu}_{\mathbf{r}}$, the tracking system of the UAV initially aims to determine $\boldsymbol{\mu}_{\boldsymbol{\omega}}$ such that the beam line intersects with the center of the receiver lens, i.e., the origin, such that the lens collects the maximum power. This leads~to
\iftoggle{OneColumn}{%
\begin{IEEEeqnarray}{lll} \label{Eq:Angles}
\mu_{\theta}=\begin{cases}
\pi+\tan^{-1}\big(\frac{\mu_y}{\mu_x}\big)&\mathrm{if}\,\,\mu_x>0 \\
\tan^{-1}\big(\frac{\mu_y}{\mu_x}\big)&\mathrm{otherwise},
\end{cases} 
\quad\text{and}\quad
\mu_{\phi}=\pi-\cos^{-1}\bigg(\frac{\mu_z}{\sqrt{\mu_x^2+\mu_y^2+\mu_z^2}}\bigg).
\end{IEEEeqnarray}
}{%
\begin{IEEEeqnarray}{lll} \label{Eq:Angles}
\mu_{\theta}=\begin{cases}
\pi+\tan^{-1}\big(\frac{\mu_y}{\mu_x}\big)&\mathrm{if}\,\,\mu_x>0 \\
\tan^{-1}\big(\frac{\mu_y}{\mu_x}\big)&\mathrm{otherwise},
\end{cases} 
\quad\text{and}\nonumber\\
\mu_{\phi}=\pi-\cos^{-1}\bigg(\frac{\mu_z}{\sqrt{\mu_x^2+\mu_y^2+\mu_z^2}}\bigg).
\end{IEEEeqnarray}
}
In other words, $\mathbb{E}\{\mathbf{b}\}=(0,0,0)$.

\subsubsection{Second Moments of RVs} 
The second moments of $\mathbf{r}$ and $\boldsymbol{\omega}$ determine how well the UAV is able to maintain its position and orientation around the mean values $\boldsymbol{\mu}_{\mathbf{r}}$ and  $\boldsymbol{\mu}_{\boldsymbol{\omega}}$, respectively. In particular, the smaller the variances of the elements of vectors $\mathbf{r}$ and $\boldsymbol{\omega}$ are, the more stable the UAV is. Hence, we consider the variances of the position and orientation of the UAV as a measure for the stability of the UAV and subsequently evaluate the performance of the FSO fronthaul link in terms of this measure.

\subsection{Statistical GML Model for Independent Gaussian Fluctuations}

Position and orientation of a hovering UAV fluctuate around their mean values even in calm weather conditions, i.e., in the absence of wind. 
These fluctuations are the result of many factors such as inherent random air fluctuations in the atmosphere around the UAV and the internal vibrations of the UAV due to e.g. the rotation of its propellers.
 Hence, invoking the central limit theorem, we model the resulting position and orientation fluctuations of the UAV as Gaussian distributed RVs. Moreover, we assume that the fluctuations are independent.
 We note that this is inline with the independent Gaussian fluctuations assumed for derivation of the statistical model for the geometric spread and pointing error due to building sway for conventional FSO links \cite{Steve_pointing_error,Alouini_Pointing}. Fluctuations $\boldsymbol{\epsilon}_{\mathbf{r}}$ and $\boldsymbol{\epsilon}_{\boldsymbol{\omega}}$ are modeled as zero-mean Gaussian random vectors, i.e.,   
\begin{IEEEeqnarray}{lll} \label{Eq:Fluc_Ind_Gauss}
\boldsymbol{\epsilon}_{\mathbf{r}}=\boldsymbol{\epsilon}_{\mathbf{r}}^{{\mathrm{IG}}}\sim\mathcal{N}(\mathbf{0},\mathbf{Q}_{\mathbf{r}}^{{\mathrm{IG}}})\quad\text{and}\quad\boldsymbol{\epsilon}_{\boldsymbol{\omega}}=\boldsymbol{\epsilon}_{\boldsymbol{\omega}}^{{\mathrm{IG}}}\sim\mathcal{N}(\mathbf{0},\mathbf{Q}_{\boldsymbol{\omega}}^{\mathrm{IG}}),\qquad
\end{IEEEeqnarray}
 where the elements of $\boldsymbol{\epsilon}_{\mathbf{r}}^{{\mathrm{IG}}}=({\epsilon}_{x}^{{\mathrm{IG}}},{\epsilon}_{y}^{{\mathrm{IG}}},{\epsilon}_{z}^{{\mathrm{IG}}})$ and $\boldsymbol{\epsilon}^{{\mathrm{IG}}}_{\boldsymbol{\omega}}=({\epsilon}_{\theta}^{{\mathrm{IG}}},{\epsilon}_{\phi}^{{\mathrm{IG}}})$ are independent Gaussian RVs. Moreover, covarinace matrices $\mathbf{Q}_{\mathbf{r}}^{\mathrm{IG}}$ and $\mathbf{Q}_{\boldsymbol{\omega}}^{\mathrm{IG}}$ are defined as $\mathbf{Q}_{\mathbf{r}}^{\mathrm{IG}}=\mathrm{diag}\{\sigma_x^2,\sigma_y^2,\sigma_z^2\}$ and
$\mathbf{Q}_{\boldsymbol{\omega}}^{\mathrm{IG}}=\mathrm{diag}\{\sigma_{\theta}^2,\sigma_{\phi}^2\}$, where $\sigma^2_{s},\,\,s\,\in\{x,y,z,\theta,\phi\},$ is the variance of the fluctuation of component $s$.  

The PDF of the GML can be derived by combining (\ref{Eq:hg_mean})-(\ref{Eq:Fluc_Ind_Gauss}). Note that in (\ref{Eq:hg_mean}), $A_0$, $t$, and $u$ are RVs since $A_0$ and $t$ depend on RV $\boldsymbol{\omega}$, and $u$ depends on both RVs $\mathbf{r}$ and $\boldsymbol{\omega}$. However, the variances of $A_0$ and $t$ are several orders of magnitude smaller than the variance of $u$. 
The reason for this is that a small variation in $\boldsymbol{\omega}$, e.g., on the order of mrad, has a significant impact on $u=\sqrt{b_y^2+b_z^2}$ since the impact of this variation on $b_y$ and $b_z$ in (\ref{Eq:FootPrint_Center}) is scaled by $r_x$ which typically has a comparatively large value (on the order of several hundred meters). On the other hand, the impact of variations in $\boldsymbol{\omega}$ on $A_0$ and $t$ is not scaled by $r_x$. 
Therefore, the fluctuations of $h_g(\mathbf{r},\boldsymbol{\omega})$ are mainly caused by the variations of the misalignment, $u$. Hence, in the following, we assume that the values of $A_0$ and $t$ are approximately constant and obtained for the average values of the position and orientation of the UAV, i.e., $\boldsymbol{\mu}_{\mathbf{r}}$ and $\boldsymbol{\mu}_{\boldsymbol{\omega}}$. In Section VI, we confirm this assumption via simulations, cf. Figs.~\ref{Fig:PDF_s}-\ref{Fig:Rate}. In addition, as shown in Appendix~\ref{App:Theo_PDF_r}, $u$ follows a Nakagami-q (Hoyt) distribution. Based on \eqref{Eq:hg_mean}, the relationship between the PDF of $h_g$ and $u$, denoted by $f_{h_g}(\cdot)$ and $f_u(\cdot)$, respectively, is~given~by
\begin{IEEEeqnarray}{lll}\label{Eq:PDF_h_PDF_u}
	f_{h_g}(h)=\frac{\sqrt{tw_L^2}}{2\sqrt{2}h\sqrt{\ln\left(\frac{A_0}{h}\right)}}f_u\left(\sqrt{\frac{tw_L^2}{2}\ln\left(\frac{A_0}{h}\right)}\right).
\end{IEEEeqnarray}
 In the following theorem, we derive the distribution of $h_g$ for small $\sigma^2_{s},\,\,s\,\in\{x,y,z,\theta,\phi\}$.

\begin{theo}\label{Theo:PDF_r}
Assuming $\sigma_s\to 0,\,s\in\{x,y,z,\theta,\phi\}$, the PDF of $h_g$ is given by
\iftoggle{OneColumn}{%
\begin{IEEEeqnarray}{lll} \label{Eq:PDF_h_In}
	f_{h_g}(h) = & \frac{\varpi }{A_0}
	\left(\frac{h}{A_0}\right)^{\frac{(1+q^2)\varpi}{2q}-1}  I_0\left(-\frac{(1-q^2)\varpi}{2q}\ln\left(\frac{h}{A_0}\right)\right), \quad 0< h \leq A_0,\quad 
	\quad
\end{IEEEeqnarray}
}{%
\begin{IEEEeqnarray}{lll} \label{Eq:PDF_h_In}
	f_{h_g}(h) =  &\frac{\varpi }{A_0}
	\left(\frac{h}{A_0}\right)^{\frac{(1+q^2)\varpi}{2q}-1}\times\nonumber\\
	  &I_0\left(-\frac{(1-q^2)\varpi}{2q}\ln\left(\frac{h}{A_0}\right)\right),
	 \quad 0< h \leq A_0,\quad 
	\quad
\end{IEEEeqnarray}
}
where $\varpi = \frac{(1+q^2)tw^2_L}{4q\Omega}$ is a constant and $I_0(\cdot)$ is the zero-order modified Bessel function of the first kind. Moreover, $q=\sqrt{\frac{\min\{\lambda_1,\lambda_2\}}{\max\{\lambda_1,\lambda_2\}}}$ and $\Omega=\lambda_1+\lambda_2$, where $\lambda_1$ and $\lambda_2$ are the eigenvalues of matrix $\boldsymbol{\Sigma}_{\mathrm{IG}}$, which is given by 
\iftoggle{OneColumn}{%
\begin{IEEEeqnarray}{lll} \label{Eq:Cov}
\boldsymbol{\Sigma}_{\mathrm{IG}} =
\begin{bmatrix}
\sigma^2_y+c_1^2\sigma^2_x+c_2^2\sigma^2_{\theta} 
& c_1c_5\sigma^2_x + c_2c_4\sigma^2_{\theta}\\
c_1c_5\sigma^2_x + c_2c_4\sigma^2_{\theta}
&  \sigma^2_z+c_3^2\sigma^2_{\phi}+c_4^2\sigma^2_{\theta}+c_5^2\sigma^2_x
\end{bmatrix}.\quad
\end{IEEEeqnarray}
}{%
\begin{IEEEeqnarray}{lll} \label{Eq:Cov}
\!\!\!\boldsymbol{\Sigma}_{\mathrm{IG}} =
\begin{bmatrix}
\sigma^2_y+c_1^2\sigma^2_x+c_2^2\sigma^2_{\theta} 
& c_1c_5\sigma^2_x + c_2c_4\sigma^2_{\theta}\\
c_1c_5\sigma^2_x + c_2c_4\sigma^2_{\theta}
&  \sigma^2_z+c_3^2\sigma^2_{\phi}+c_4^2\sigma^2_{\theta}+c_5^2\sigma^2_x
\end{bmatrix}\!\!.\quad
\end{IEEEeqnarray}
}
In \eqref{Eq:Cov}, $c_1=-\tan\mu_{\theta}$, $c_2=-\frac{\mu_x}{\cos^2\mu_{\theta}}$, $c_3=\frac{\mu_x}{\sin^2\mu_{\phi}\cos\mu_{\theta}}$, $c_4=-\frac{\mu_x\cot\mu_{\phi}\tan\mu_{\theta}}{\cos\mu_{\theta}}$, and $c_5=-\frac{\cot\mu_{\phi}}{\cos\mu_{\theta}}$ are constants.
\end{theo}
\begin{IEEEproof}
The proof is given in Appendix~\ref{App:Theo_PDF_r}.
\end{IEEEproof}
Note that the PDF of $f_{h_g}(h)$ in (\ref{Eq:PDF_h_In}) has an indeterminate form at $h=0$. Its value can be found for $q\neq 1$ as
\iftoggle{OneColumn}{%
\begin{IEEEeqnarray}{lll} \label{Eq:PDFat0}
\underset{h\to0}{\lim} \,\, f_{h_g}(h) \overset{(a)}{=}\underset{h\to0}{\lim}\,\frac{\sqrt{q\varpi }}{\sqrt{\pi(1-q^2)}A_0}
\left[\ln\left(\frac{A_0}{h}\right) \right]^{-\frac{1}{2}}\left(\frac{h}{A_0}\right)^{q\varpi -1} =\begin{cases}
0,&q\varpi \geq 1\\
\infty, &q\varpi<1,
\end{cases}
\end{IEEEeqnarray}
}{%
\begin{IEEEeqnarray}{lll} \label{Eq:PDFat0}
\underset{h\to0}{\lim} \,\, f_{h_g}(h) \overset{(a)}{=}&\underset{h\to0}{\lim}\,\frac{\sqrt{q\varpi }}{\sqrt{\pi(1-q^2)}A_0}
\left[\ln\left(\frac{A_0}{h}\right) \right]^{-\frac{1}{2}}\nonumber\\
&\times\left(\frac{h}{A_0}\right)^{q\varpi -1} =\begin{cases}
0,&q\varpi \geq 1\\
\infty, &q\varpi<1,
\end{cases}
\end{IEEEeqnarray}
}
where for equality $(a)$, we used $\underset{z\to \infty}{\lim}I_0(z)=\frac{\exp(z)}{\sqrt{2\pi z}}$ \cite[Eq.~(9.7.1)]{Handbook}. In fact, (\ref{Eq:PDFat0}) shows that for a wider beam and smaller variances of the fluctuations, for which $q\varpi\geq1$ is met, the channel quality becomes better since the probability of small channel coefficient values approaches zero. On the other hand, having a wide beam reduces the maximum fraction of power collected by the receiver lens, $A_0$, cf. (\ref{Eq:A_0}). Therefore, there is a trade-off between  $A_0$ and $q\varpi$ when choosing the beam width (beam divergence angle).  In the following corollary, we investigate the special case when the beam is perpendicular w.r.t. the lens plane.

\begin{corol}\label{Corol:uRayl} 
When the laser beam is perpendicular w.r.t. the lens plane, i.e., $\mu_y=\mu_z=0$, $\mu_{\theta}=\pi$, and $\mu_{\phi}=\pi/2$, $\boldsymbol{\Sigma}_{\mathrm{IG}}$ is given by
\begin{IEEEeqnarray}{lll} \label{Eq:Cov_Rayleigh}
\boldsymbol{\Sigma}_{\mathrm{IG}} =
\begin{bmatrix}
\sigma^2_y+\mu_{x}^2\sigma^2_{\theta}  & 0\\
0&  \sigma^2_z+\mu_x^2\sigma^2_{\phi}
\end{bmatrix},\quad
\end{IEEEeqnarray}
which has eigenvalues $\lambda_1=\sigma^2_y+\mu_{x}^2\sigma^2_{\theta}$ and $\lambda_2=\sigma^2_z+\mu_x^2\sigma^2_{\phi}$. Hereby, assuming $\sigma^2_y=\sigma^2_z\triangleq\sigma^2_{p}$ and $\sigma^2_{\theta} =\sigma^2_{\phi}\triangleq\sigma^2_{o}$ leads to $q=1$ and RV $u$ is Rayleigh distributed \cite{Steve_pointing_error}. Therefore, the PDF of $h_g$ simplifies to
\begin{IEEEeqnarray}{lll} \label{Eq:PDF_Rayleigh}
f_{h_g}(h) = & \frac{\varpi}{A_0}
\left(\frac{h}{A_0}\right)^{\varpi-1}, \quad 0\leq h \leq A_0,\quad 
\end{IEEEeqnarray}
where $\varpi=\frac{tw^2_L}{4(\sigma^2_p+\mu^2_x\sigma^2_o)}$. 
\end{corol}
\begin{IEEEproof}
The proof follows by substitution of $q=1$ and $\Omega=2(\sigma^2_p+\mu^2_x\sigma^2_o)$ into (\ref{Eq:PDF_h_In}).
\end{IEEEproof}

Depending on the value of $\varpi$, the PDF of the GML in (\ref{Eq:PDF_Rayleigh}) shows the following behavior. \textit{i)} If $\varpi> 1$ holds, the probability of small channel coefficients becomes very small, i.e., $\underset{h\to 0}{\lim}\,f_{h_g}(h)=0$. As a special case when the UAV is fully stable, i.e., $\sigma_p=\sigma_o=0$ leading to $\varpi\to \infty$, random fluctuations are not present anymore and the GML becomes a deterministic function of the given position and orientation of the UAV. In other words, the PDF of the GML becomes a Dirac function at $A_0$, i.e., $f_{h_g}(h)=\delta(h-A_0)$. \textit{ii)} If $\varpi=1$ holds, the GML is uniformly distributed in $[0,A_0]$, i.e., $f_{h_g}(h) = \frac{1}{A_0}$. \textit{iii)} If $\varpi< 1$ holds, the channel quality deteriorates and the probability of small channel coefficients becomes very large, i.e., $\underset{h\to 0}{\lim}\,f_{h_g}(h)=\infty$. Recall that for the non-orthogonal case in (\ref{Eq:PDFat0}), when $q\neq 1$, we have two cases $\underset{h\to 0}{\lim}\,f_{h_g}(h)\in\{0,\infty\}$ depending on the value of $q\varpi$; whereas for the orthogonal case, $q=1$, we have three cases $\underset{h\to 0}{\lim}\,f_{h_g}(h)\in\{0,\frac{1}{A_0},\infty\}$ depending on the value of $\varpi$.

The simplified matrix $\boldsymbol{\Sigma}_{\mathrm{IG}}$ in (\ref{Eq:Cov_Rayleigh}) reveals that the GML is much more sensitive to the variance of the orientation, $\sigma^2_{o}$, than to the variance of the position, $\sigma^2_{p}$, since $\sigma^2_{o}$ is scaled by the average distance between the UAV and the CU, i.e., $\mathbb{E}\{\|\mathbf{r}\|\}=\mu_x$. Another interesting observation from (\ref{Eq:Cov_Rayleigh}) is that the variation of the position and orientation of the UAV along the $x$ axis does not affect the GML since $\sigma_x^2$ does not appear in (\ref{Eq:Cov_Rayleigh}). The reason for this is that, since the beam is orthogonal to the receiver lens plane, the optical beam propagates along the $x$ axis, and therefore, small changes of the position of the UAV in $x$ direction do not affect the power collected by the PD. Finally, we note that despite the presence of both position and orientation fluctuations, (\ref{Eq:PDF_Rayleigh}) has a similar form as the geometric spread and the pointing error in conventional FSO systems where only position fluctuations are present \cite{Steve_pointing_error,Alouini_Pointing}.


\subsection{Statistical GML Model for Correlated Gaussian Fluctuations}
Now, we consider the case where there is a weak wind along a specific direction denoted by $\mathbf{v}=(v_x,v_y,v_z)$. In this scenario, it is expected that the wind causes larger fluctuations of RV $\mathbf{r}$ along the direction of $\mathbf{v}$. Similarly, depending on the geometry of the UAV\footnote{For a perfect spherical object, due to symmetry, the  force applied on its surface by wind does not create rotational forces. However, for practical UAV geometries, the impact of the wind force will be dominant in a certain direction which causes rotational forces. The exact direction of the rotational force depends on the object geometry and the direction of the wind and cannot be specified a priori.}, the wind may cause larger fluctuations of $\boldsymbol{\omega}$ in a certain direction, denoted by $\boldsymbol{\tau}=(\tau_{\theta},\tau_{\phi})$. Hence, we model the fluctuations of  $\mathbf{r}$ and  $\boldsymbol{\omega}$ as \textit{correlated} Gaussian RVs. Note that the total fluctuations are the result of both independent and correlated Gaussian distributed variations. In particular, the fluctuations are modeled as
\iftoggle{OneColumn}{%
\begin{IEEEeqnarray}{lll} \label{Eq:Gaussian_Correlated}
\boldsymbol{\epsilon}_{\mathbf{r}}=\boldsymbol{\epsilon}_{\mathbf{r}}^{{\mathrm{IG}}}+\boldsymbol{\epsilon}_{\mathbf{r}}^{{\mathrm{CG}}}\sim\mathcal{N}(\mathbf{0},\mathbf{Q}_{\mathbf{r}}^{{\mathrm{IG}}}+\mathbf{Q}_{\mathbf{r}}^{{\mathrm{CG}}}) \quad \text{and} \quad
\boldsymbol{\epsilon}_{\boldsymbol{\omega}}=\boldsymbol{\epsilon}_{\boldsymbol{\omega}}^{{\mathrm{IG}}}+\boldsymbol{\epsilon}_{\boldsymbol{\omega}}^{{\mathrm{CG}}}\sim\mathcal{N}(\mathbf{0},\mathbf{Q}_{\boldsymbol{\omega}}^{{\mathrm{IG}}}+\mathbf{Q}_{\boldsymbol{\omega}}^{{\mathrm{CG}}}).
\end{IEEEeqnarray}
}{%
\begin{IEEEeqnarray}{lll} \label{Eq:Gaussian_Correlated}
\boldsymbol{\epsilon}_{\mathbf{r}}=\boldsymbol{\epsilon}_{\mathbf{r}}^{{\mathrm{IG}}}+\boldsymbol{\epsilon}_{\mathbf{r}}^{{\mathrm{CG}}}\sim\mathcal{N}(\mathbf{0},\mathbf{Q}_{\mathbf{r}}^{{\mathrm{IG}}}+\mathbf{Q}_{\mathbf{r}}^{{\mathrm{CG}}}) \quad \text{and}\nonumber\\
\boldsymbol{\epsilon}_{\boldsymbol{\omega}}=\boldsymbol{\epsilon}_{\boldsymbol{\omega}}^{{\mathrm{IG}}}+\boldsymbol{\epsilon}_{\boldsymbol{\omega}}^{{\mathrm{CG}}}\sim\mathcal{N}(\mathbf{0},\mathbf{Q}_{\boldsymbol{\omega}}^{{\mathrm{IG}}}+\mathbf{Q}_{\boldsymbol{\omega}}^{{\mathrm{CG}}}).
\end{IEEEeqnarray}
}
 Here, $\boldsymbol{\epsilon}_{\mathbf{r}}^{{\mathrm{CG}}}=(\epsilon_{x}^{{\mathrm{CG}}},\epsilon_{y}^{{\mathrm{CG}}},\epsilon_{z}^{{\mathrm{CG}}})$ denotes a random vector with correlated Gaussian distributed elements that models the fluctuation of the position of the UAV due to the wind. In particular, we model $\boldsymbol{\epsilon}_{\mathbf{r}}^{{\mathrm{CG}}}$ as $\boldsymbol{\epsilon}_{\mathbf{r}}^{{\mathrm{CG}}}=\delta^{\mathrm{G}}\mathbf{v}$, where $\delta^{\mathrm{G}}\sim\mathcal{N}(0,\zeta^2)$ denotes a zero-mean normal RV with variance $\zeta^2$. Moreover, $\mathbf{Q}_{\mathbf{r}}^{{\mathrm{CG}}}=\zeta^2\mathbf{v}^{\mathsf{T}}\mathbf{v}$ is the covariance matrix of $\boldsymbol{\epsilon}_{\mathbf{r}}^{{\mathrm{CG}}}$.  Similarly, we have $\boldsymbol{\epsilon}_{\boldsymbol{\omega}}^{{\mathrm{CG}}}=(\epsilon_{\theta}^{{\mathrm{CG}}},\epsilon_{\phi}^{{\mathrm{CG}}})=\delta^{\mathrm{G}}\boldsymbol{\tau}$ and its covariance matrix is given by $\mathbf{Q}_{\boldsymbol{\omega}}^{{\mathrm{CG}}}=\zeta^2\boldsymbol{\tau}^{\mathsf{T}}\boldsymbol{\tau}$~\footnote{For notational consistency, we assume that the unit of $\delta^{\mathrm{G}}$ is meter. This implies that $\mathbf{v}$ is unitless and the unit of $\boldsymbol{\tau}$ is rad.m$^{-1}$.}. We note that similar to the independent Gaussian fluctuation model, for the correlated Gaussian fluctuation model, $u$ follows a Nakagami-q (Hoyt) distribution, cf. Appendix~\ref{App:Theo_PDF_u_Gcorrelated}. In the following theorem, we derive the PDF of $h_g$.


\begin{theo}\label{Theo:PDF_u_Gcorrelated}
Assuming $\sigma_s\to 0,\,s\in\{x,y,z,\theta,\phi\}$ and $\zeta\to 0$, $h_g$ follows the PDF in (\ref{Eq:PDF_h_In}) if matrix $\boldsymbol{\Sigma}_{\mathrm{IG}}$ is replaced by $\boldsymbol{\Sigma}_{\mathrm{T}}=\boldsymbol{\Sigma}_{\mathrm{IG}}+\boldsymbol{\Sigma}_{\mathrm{CG}}$, where $\boldsymbol{\Sigma}_{\mathrm{CG}}$ is given by
\begin{IEEEeqnarray}{lll} \label{Eq:Cov_Gcorrelated}
	\boldsymbol{\Sigma}_{\mathrm{CG}}=\zeta^2
\begin{bmatrix}
c_6^2 
& c_6c_7 \\
c_6c_7 
&  c_7^2
\end{bmatrix}.\quad
\end{IEEEeqnarray}
Here, $c_6=v_y+v_xc_1+\tau_{\theta}c_2$,  $c_7=v_z+v_xc_5+\tau_{\phi}c_3+\tau_{\theta}c_4$, and constants $c_1$-$c_5$ are defined in Theorem~\ref{Theo:PDF_r}.
\end{theo}
\begin{IEEEproof}
The proof is given in Appendix~\ref{App:Theo_PDF_u_Gcorrelated}.
\end{IEEEproof}

 In the following, we consider the special case where the impact of the wind on the fluctuations of $\mathbf{r}$ and $\boldsymbol{\omega}$ is dominant, i.e., $\mathrm{Tr}\{\boldsymbol{\Sigma}_{\mathrm{CG}}\}\geq \mathrm{Tr}\{\boldsymbol{\Sigma}_{\mathrm{IG}}\}$. In this case, $u$ follows a one-sided Gaussian distribution given in Appendix~\ref{App:Corol_Gauss_Single}.
\begin{corol}\label{corol:halfnormal}
	 For the special case, where the impact of wind is dominant, the PDF of $h_g$ simplifies to
	 \iftoggle{OneColumn}{%
	\begin{IEEEeqnarray}{lll} \label{Eq:PDF_halfnormal}
		f_{h_g}(h) = & \frac{\sqrt{\varpi}}{\sqrt{\pi}A_0}\left[\ln\left(\frac{A_0}{h}\right)\right]^{-\frac{1}{2}}
		\left(\frac{h}{A_0}\right)^{\varpi-1}, \quad 0\leq h \leq A_0,\quad 
	\end{IEEEeqnarray}
}{%
	\begin{IEEEeqnarray}{lll} \label{Eq:PDF_halfnormal}
		f_{h_g}(h) = & \frac{\sqrt{\varpi}}{\sqrt{\pi}A_0}\left[\ln\left(\frac{A_0}{h}\right)\right]^{-\frac{1}{2}}
		\left(\frac{h}{A_0}\right)^{\varpi-1}\hspace{-4mm},  0\leq h \leq A_0,\quad\,\
	\end{IEEEeqnarray}
}
	where $\varpi=\frac{tw^2_L}{4\zeta^2(c_6^2+c_7^2)}$. 
\end{corol}
\begin{IEEEproof}
The proof is given in Appendix~\ref{App:Corol_Gauss_Single}.
\end{IEEEproof}

 For small channel coefficients, i.e., $h\to 0$, the PDF in \eqref{Eq:PDF_halfnormal} has the following transient behavior: If $\varpi\geq 1$, $\underset{h\to 0}{\lim}f_{h_g}(h)=0$ and if $\varpi< 1$, $\underset{h\to 0}{\lim}f_{h_g}(h)=\infty$ which is similar to the behavior in the independent Gaussian case, cf. (\ref{Eq:PDFat0}). Moreover, $\underset{h\to A_0}{\lim}f_{h_g}(h)=\infty$ holds which is different from the independent Gaussian and general correlated Gaussian cases where $\underset{h\to A_0}{\lim}f_{h_g}(h)=\frac{\varpi}{A_0}$ is bounded for $q\neq 0$, cf. (\ref{Eq:PDF_h_In}). 

\subsection{Statistical GML Model for Correlated Uniform Fluctuations}
If strong wind is present, the fluctuations of the position and orientation of the UAV are relatively large compared to those for calm and weakly windy weather conditions. On the other hand, for practical UAVs, it is reasonable to assume that despite being large, the fluctuations are bounded. In this case, assuming a Gaussian distribution for the fluctuations is not appropriate. Instead, the uniform distribution is a better model for the fluctuations of the position and orientation of the UAV. Note that in the absence of prior knowledge, the uniform distribution is a widely-adopted choice for bounded RVs, see e.g., the application of the uniform distribution for robustness analysis in \cite{Uniform1}, uncertainty analysis in \cite{Uniform2}, and worst-case analysis in \cite{Uniform3}. Similar to Section IV-C, we assume that the UAV position and orientation fluctuations are stronger in a certain direction which may be caused by e.g. wind. Furthermore, we assume that the effect of the wind is the dominant source of the fluctuations. More specifically, the fluctuations of $\mathbf{r}$ and $\boldsymbol{\omega}$ are modeled as 
\begin{IEEEeqnarray}{lll} \label{Eq:Uniform_r_w}
\boldsymbol{\epsilon}_{\mathbf{r}}= \boldsymbol{\epsilon}_{\mathbf{r}}^{\mathrm{CU}}\quad \text{and} \quad
\boldsymbol{\epsilon}_{\boldsymbol{\omega}}=\boldsymbol{\epsilon}_{\boldsymbol{\omega}}^{\mathrm{CU}},
\end{IEEEeqnarray}
where $\boldsymbol{\epsilon}_{\mathbf{r}}^{\mathrm{CU}}=(\epsilon_{x}^{{\mathrm{CU}}},\epsilon_{y}^{{\mathrm{CU}}},\epsilon_{z}^{{\mathrm{CU}}})$ denotes the fluctuation of position of the UAV due to strong wind and is given by $\boldsymbol{\epsilon}_{\mathbf{r}}^{\mathrm{CU}}=\delta^{\mathrm{U}}\mathbf{v}$, where $\delta^{\mathrm{U}}\sim\mathcal{U}\left(-\sqrt{3}\xi,\sqrt{3}\xi\right)$ is a zero-mean uniformly distributed RV with variance $\xi$. Similarly, $\boldsymbol{\epsilon}_{\boldsymbol{\omega}}^{\mathrm{CU}}=(\epsilon_{\theta}^{{\mathrm{CU}}},\epsilon_{\phi}^{{\mathrm{CU}}})=\delta^{\mathrm{U}}\boldsymbol{\tau}$ denotes the fluctuation of the orientation of the UAV caused by wind in a specific direction $\boldsymbol{\tau}$. In this case, the misalignment, $u$, follows a uniform distribution, $u\sim\mathcal{U}\left(0,\sqrt{3(c_6^2+c_7^2)}\,\xi\right)$, cf. Appendix~\ref{App:Theo_PDF_r_Uniform}. The following theorem provides the PDF of the GML for correlated uniform fluctuations.

\begin{theo}\label{Theo:PDF_u2_Uniform}
Assuming $\xi\to 0$, the PDF of  $h_g$ is given by
\begin{IEEEeqnarray}{lll}\label{Eq:PDF_h_Unif}
f_{h_g}(h)=\frac{\alpha_1}{h\sqrt{\ln\left(\frac{A_0}{h}\right)}}, \quad h_1\leq h \leq A_0,
\end{IEEEeqnarray}
where $\alpha_1=\sqrt{\frac{tw^2_L}{24(c_6^2+c_7^2)\xi^2}}$ and $h_1=A_0\exp\left(-\frac{6(c_6^2+c_7^2)\xi^2}{tw^2_L}\right)$. 
\end{theo}
\begin{IEEEproof}
The proof is given in Appendix~\ref{App:Theo_PDF_r_Uniform}.
\end{IEEEproof}
We note that PDF $f_{h_g}(h)$ assumes large values at $h=A_0$, i.e., $\underset{h\to A_0}{\lim}f_{h_g}(h)=\infty$.

\section{Performance Analysis}
In this section, we analyze the outage probability and ergodic rate of the considered UAV-based FSO system.

\subsection{Outage Probability}
The outage probability is defined as the probability that the SNR, denoted by $\gamma$, falls below a predefined threshold, $\gamma_{\mathrm{thr}}$. For the channel model in (\ref{Eq:signal}), the SNR is defined as $\gamma=\eta^2h_p^2h_g^2\bar{\gamma}$ where $\bar{\gamma}=\frac{P^2}{\sigma^2_n}$ is the transmit SNR. Therefore, the outage probability is obtained~as a function of the cumulative distribution function (CDF) of the GML as follows
\iftoggle{OneColumn}{%
\begin{IEEEeqnarray}{lll} \label{Eq:out}
P_{\mathrm{out}}=\mathrm{Pr}\{\gamma\leq\gamma_{\mathrm{thr}}\}=\mathrm{Pr}\left\{h_g\leq \frac{\sqrt{{\gamma_{\mathrm{thr}}}}}{\eta h_p\sqrt{\bar{\gamma}}}\right\}=F_{h_g}\left(\frac{\sqrt{{\gamma_{\mathrm{thr}}}}}{\eta h_p\sqrt{\bar{\gamma}}}\right),\quad 0 \leq \frac{\sqrt{{\gamma_{\mathrm{thr}}}}}{\eta h_p\sqrt{\bar{\gamma}}}\leq A_0,\quad
\end{IEEEeqnarray}
}{%
\begin{IEEEeqnarray}{lll} \label{Eq:out}
P_{\mathrm{out}}&=\mathrm{Pr}\{\gamma\leq\gamma_{\mathrm{thr}}\}=\mathrm{Pr}\left\{h_g\leq \frac{\sqrt{{\gamma_{\mathrm{thr}}}}}{\eta h_p\sqrt{\bar{\gamma}}}\right\}\nonumber\\
&=F_{h_g}\left(\frac{\sqrt{{\gamma_{\mathrm{thr}}}}}{\eta h_p\sqrt{\bar{\gamma}}}\right),\quad 0 \leq \frac{\sqrt{{\gamma_{\mathrm{thr}}}}}{\eta h_p\sqrt{\bar{\gamma}}}\leq A_0,\quad
\end{IEEEeqnarray}
}
where $F_{h_g}(\cdot)$ denotes the CDF of the GML.
In the following, we derive the outage probability for different fluctuation scenarios.

\subsubsection{Independent Gaussian Fluctuations}
For the case of independent Gaussian fluctuations, using (\ref{Eq:PDF_h_In}),  $P_{\mathrm{out}}$ can be written as 
\begin{IEEEeqnarray}{lll} \label{Eq:CDF_h}
	P_{\mathrm{out}} =
	 1-Q\left(a,b\right)+Q\left(b,a\right),
\end{IEEEeqnarray}
for  $0 \leq \frac{\sqrt{{\gamma_{\mathrm{thr}}}}}{\eta h_p\sqrt{\bar{\gamma}}}\leq A_0$, where $a=\frac{1+q}{2q}g$,  $b=\frac{1-q}{2q}g$, and $g=\sqrt{\frac{1+q^2}{\Omega}\frac{tw^2_L}{2}\ln\left(\frac{\eta h_pA_0\sqrt{\bar{\gamma}}}{\sqrt{{\gamma_{\mathrm{thr}}}}}\right)}$ \cite{Nakagami_Hoyt}. In the following, we simplify (\ref{Eq:CDF_h}) for some special cases.


\begin{corol}
For independent Gaussian fluctuations, if the beam is orthogonal to the lens plane,  $P_{\mathrm{out}}$ can be obtained as
 \begin{IEEEeqnarray}{lll} \label{Eq:CDF_Rayleigh}
 	P_{\mathrm{out}} = & 	\left(\frac{\sqrt{{\gamma_{\mathrm{thr}}}}}{\eta h_pA_0\sqrt{\bar{\gamma}}}\right)^{\varpi}, \quad 0\leq x \leq A_0.
 \end{IEEEeqnarray}

\end{corol}
\begin{IEEEproof}
	After integrating the PDF of the GML for this special case (see \eqref{Eq:PDF_Rayleigh}), the CDF of the GML and hence, $P_{\mathrm{out}}$ can be obtained as in \eqref{Eq:CDF_Rayleigh}.
\end{IEEEproof}

Eq. \eqref{Eq:CDF_Rayleigh} reveals that the diversity gain of the FSO link is
\begin{IEEEeqnarray}{lll}
d=-\underset{\bar{\gamma}\to \infty}{\lim}\frac{\log(P_{\mathrm{out}})}{\log(\bar{\gamma})}=\frac{\varpi}{2}=\frac{tw^2_L}{8(\sigma^2_p+\mu^2_x\sigma^2_o)}.
\end{IEEEeqnarray}

\begin{corol}\label{Corol:Asymp_Outage}
For high SNRs, i.e., for large values of arguments $a$ and $b$, (\ref{Eq:CDF_h}) can be simplified~as~\cite{MarcumQ}
\iftoggle{OneColumn}{%
\begin{IEEEeqnarray}{lll}\label{Eq:Corollary_outage}
\underset{\bar{\gamma}\to \infty}{\lim} P_{\mathrm{out}}=\underset{\bar{\gamma}\to \infty}{\lim} \left(\sqrt{\frac{a}{b}}+\sqrt{\frac{b}{a}}\right)Q\left(a-b\right)= a_t\left(\frac{1}{\bar{\gamma}}\right)^{\frac{(1+q^2)tw_L^2}{8\Omega}}\left[\ln\left(\frac{\bar{\gamma}}{b_t^2}\right)\right]^{-\frac{1}{2}},
\end{IEEEeqnarray}
}{%
\begin{IEEEeqnarray}{lll}\label{Eq:Corollary_outage}
\underset{\bar{\gamma}\to \infty}{\lim} P_{\mathrm{out}}&=\underset{\bar{\gamma}\to \infty}{\lim} \left(\sqrt{\frac{a}{b}}+\sqrt{\frac{b}{a}}\right)Q\left(a-b\right)\nonumber\\
&= a_t\left(\frac{1}{\bar{\gamma}}\right)^{\frac{(1+q^2)tw_L^2}{8\Omega}}\left[\ln\left(\frac{\bar{\gamma}}{b_t^2}\right)\right]^{-\frac{1}{2}},
\end{IEEEeqnarray}
}
where $a_t=\frac{2\sqrt{2\Omega}b_t^{\frac{(1+q^2)tw_L^2}{4\Omega}}}{\sqrt{\pi(1-q^4)tw_L^2}}$ and $b_t=\frac{\sqrt{{\gamma_{\mathrm{thr}}}}}{\eta h_pA_0}$.
\end{corol}
\begin{IEEEproof}
	The proof is given in Appendix~\ref{App:Corollary}.
\end{IEEEproof}

Based on (\ref{Eq:Corollary_outage}), the diversity gain is
\begin{IEEEeqnarray}{lll}\label{Eq:Div_IG}
d =\frac{(1+q^2)tw_L^2}{8\Omega}
\end{IEEEeqnarray}
 since, as $\bar{\gamma}\to \infty$, the impact of the logarithmic term $\left[\ln\left(\frac{\bar{\gamma}}{b_t^2}\right)\right]^{-\frac{1}{2}}$ in (\ref{Eq:Corollary_outage}) becomes negligible compared to that of the polynomial term $\left(\frac{1}{\bar{\gamma}}\right)^{\frac{(1+q^2)tw_L^2}{8\Omega}}$.

\subsubsection{Correlated Gaussian Fluctuations}
In this case, if both independent and correlated Gaussian fluctuations are present, then the same expression for the outage probability holds, as for the independent Gaussian scenario, cf. \eqref{Eq:CDF_h}-\eqref{Eq:Div_IG}. If the impact of wind on the fluctuations is dominant, we obtain
\begin{IEEEeqnarray}{lll}\label{Eq:CrlGauss_outage}
 P_{\mathrm{out}} =2Q\left(\sqrt{\frac{tw^2_L}{2\zeta^2(c_6^2+c_7^2)}\ln\left(\frac{\eta h_pA_0\sqrt{\bar{\gamma}}}{\sqrt{{\gamma_{\mathrm{thr}}}}}\right)}\right).
\end{IEEEeqnarray}


Based on (\ref{Eq:CrlGauss_outage}) and using the same approximation for the Gaussian Q-function as in Appendix~\ref{App:Corollary}, the diversity gain can be obtained as 
\begin{IEEEeqnarray}{lll}
d =\frac{tw_L^2}{8\zeta^2(c_6^2+c_7^2)}.
\end{IEEEeqnarray}
\subsubsection{Correlated Uniform Fluctuations}
For uniform distributed fluctuations,
\begin{IEEEeqnarray}{lll}\label{Eq:Unif_outage}
 P_{\mathrm{out}} =1-\frac{\sqrt{tw_L^2\ln\left(\frac{\eta h_pA_0\sqrt{\bar{\gamma}}}{\sqrt{{\gamma_{\mathrm{thr}}}}}\right)}}{\sqrt{6(c_6^2+c_7^2)\xi^2}},\, h_1\leq \frac{\sqrt{{\gamma_{\mathrm{thr}}}}}{\eta h_p\sqrt{\bar{\gamma}}}\leq A_0.\quad\quad
\end{IEEEeqnarray}
 Note that, in this case, the outage probability is zero if the transmit SNR is larger than a critical value, $\bar{\gamma}_{\mathrm{crt}}$, i.e., $\bar{\gamma}\geq\bar{\gamma}_{\mathrm{crt}}$, where 
\begin{IEEEeqnarray}{lll}\label{Eq:gamma_crt} 
 \bar{\gamma}_{\mathrm{crt}}=\frac{\gamma_{\mathrm{thr}}}{\eta^2h_p^2h_1^2}. 
 \end{IEEEeqnarray}

\subsection{Ergodic Rate}
 For an IM/DD FSO channel, the capacity is not known. Nevertheless, in~\cite[Eq. 26]{FSO_Cap}, the following ergodic rate has been shown to be achievable
 \iftoggle{OneColumn}{%
\begin{IEEEeqnarray}{lll} \label{Eq:R}
	\bar{R}=\frac{1}{2}\mathbb{E}_{\gamma}\left\{\log_2\left(1+\frac{e}{2\pi}\gamma\right)\right\}=\frac{1}{2}\mathbb{E}_{h_g}\left\{\log_2\left(1+ch_g^2\right)\right\},\quad \text{bits/symbol},
\end{IEEEeqnarray}
}{%
\begin{IEEEeqnarray}{lll} \label{Eq:R}
	\bar{R}&=\frac{1}{2}\mathbb{E}_{\gamma}\left\{\log_2\left(1+\frac{e}{2\pi}\gamma\right)\right\}\nonumber\\
	&=\frac{1}{2}\mathbb{E}_{h_g}\left\{\log_2\left(1+ch_g^2\right)\right\},\quad \text{bits/symbol},
\end{IEEEeqnarray}
}
 where $c=\frac{e}{2\pi}\eta^2h_p^2\bar{\gamma}$. In the following, we analyze the ergodic rate at high SNR. In particular, for high SNR, we have
 \iftoggle{OneColumn}{%
 \begin{IEEEeqnarray}{lll} \label{Eq:R_Geo_highSNR}
 	\underset{{\bar{\gamma}\to \infty}}{\lim}\bar{R}=\frac{1}{2}\mathbb{E}_{h_g}\left\{\log_2\left(ch_g^2\right)\right\}=\overset{\bar{R}_{\max}}{\overbrace{\frac{1}{2}\log_2(cA_0^2)}}-\overset{\Delta\bar{R}_g}{\overbrace{\frac{2}{tw^2_L\ln(2)}\mathbb{E}_{u}\{u^2\}}},
 \end{IEEEeqnarray}
}{%
 \begin{IEEEeqnarray}{lll} \label{Eq:R_Geo_highSNR}
 	\underset{{\bar{\gamma}\to \infty}}{\lim}\bar{R}&=\frac{1}{2}\mathbb{E}_{h_g}\left\{\log_2\left(ch_g^2\right)\right\}\nonumber\\
 	&=\overset{\bar{R}_{\max}}{\overbrace{\frac{1}{2}\log_2(cA_0^2)}}-\overset{\Delta\bar{R}_g}{\overbrace{\frac{2}{tw^2_L\ln(2)}\mathbb{E}_{u}\{u^2\}}},
 \end{IEEEeqnarray}
}
where $\bar{R}_{\max}$ is the maximum achievable ergodic rate without misalignment, i.e., $u=0$, $\Delta\bar{R}_g$ is the loss in ergodic rate due misalignment, and $\mathbb{E}_{u}\{u^2\}$ denotes the expected value of the squared misalignment, i.e., $u^2$. Note that $\Delta\bar{R}_g$ depends on the distribution of the fluctuations but $\bar{R}_{\max}$ is independent of it and only depends on mean value of the UAV's  position and orientation, the beam width (beam divergence angle), the transmit SNR, as well as the area and responsivity of the PD. In the following, we evaluate $\Delta\bar{R}_g$ for the considered independent/correlated Gaussian and correlated uniform fluctuation models.

\begin{corol}
For the considered fluctuation models, $\Delta\bar{R}_g$ in bits/symbol is given by 
\iftoggle{OneColumn}{%
\begin{IEEEeqnarray}{lll}\label{Eq:deltaR}
\Delta\bar{R}_g=\frac{2}{tw^2_L\ln(2)} 
\times\begin{cases}
\lambda_1+\lambda_2, & \text{independent/correlated Gaussian}\\
(c_6^2+c_7^2)\zeta^2, & \text{correlated Gaussian (if the wind is dominant)}\\
(c_6^2+c_7^2)\xi^2, &\text{uniform}.
\end{cases}\quad
 \end{IEEEeqnarray}
}{%
\begin{IEEEeqnarray}{lll}\label{Eq:deltaR}
\Delta\bar{R}_g=\frac{2}{tw^2_L\ln(2)} 
\times\\
\begin{cases}
\lambda_1+\lambda_2, & \text{independent/correlated Gaussian}\\
(c_6^2+c_7^2)\zeta^2, & \text{correlated Gaussian (wind is dominant)}\\
(c_6^2+c_7^2)\xi^2, &\text{uniform}.
\end{cases}\quad\nonumber
 \end{IEEEeqnarray}
}
\end{corol}
\begin{IEEEproof}
For both independent and correlated Gaussian fluctuation models, $u$ follows a Hoyt distribution. Hence, its second moment is $\mathbb{E}_{u}\{u^2\}=\Omega$ \cite[Eq.~(2.12)]{Alouini_Digital_Comm_Hoyt} with $\Omega=\lambda_1+\lambda_2$. 
For the correlated Gaussian case when the effect of wind is dominant, i.e., $\mathrm{Tr}\{\boldsymbol{\Sigma}_{\mathrm{CG}}\}\geq \mathrm{Tr}\{\boldsymbol{\Sigma}_{\mathrm{IG}}\}$, $\Omega=\lambda_1=(c_6^2+c_7^2)\zeta^2$, since $\lambda_2=0$. 
On the other hand, for the uniform fluctuation model, $u$ follows a uniform distribution with second-order moment $\mathbb{E}_{u}\{u^2\}=(c_6^2+c_7^2)\xi^2$. Substituting  $\mathbb{E}_{u}\{u^2\}$ into $\Delta\bar{R}_g$ in (\ref{Eq:R_Geo_highSNR}) leads to  (\ref{Eq:deltaR}) and concludes the proof.
\end{IEEEproof}

As can be observed from (\ref{Eq:deltaR}), the rate loss due to misalignment, $\Delta\bar{R}_g$, depends on the stability of the UAV through variables $\lambda_1$, $\lambda_2$, $\zeta$ or $\xi$. Thereby, $\Delta\bar{R}_g$ decreases and, as a result, $\bar{R}$ increases as the UAV becomes more stable.

\section{Simulations and Discussions}\label{Sec:Sim}
\begin{figure*}[!tbp]
	\centering
	\begin{minipage}[b]{0.49\textwidth}
		\centering
		\hspace{-0.5cm}
		\includegraphics[width=1\linewidth]{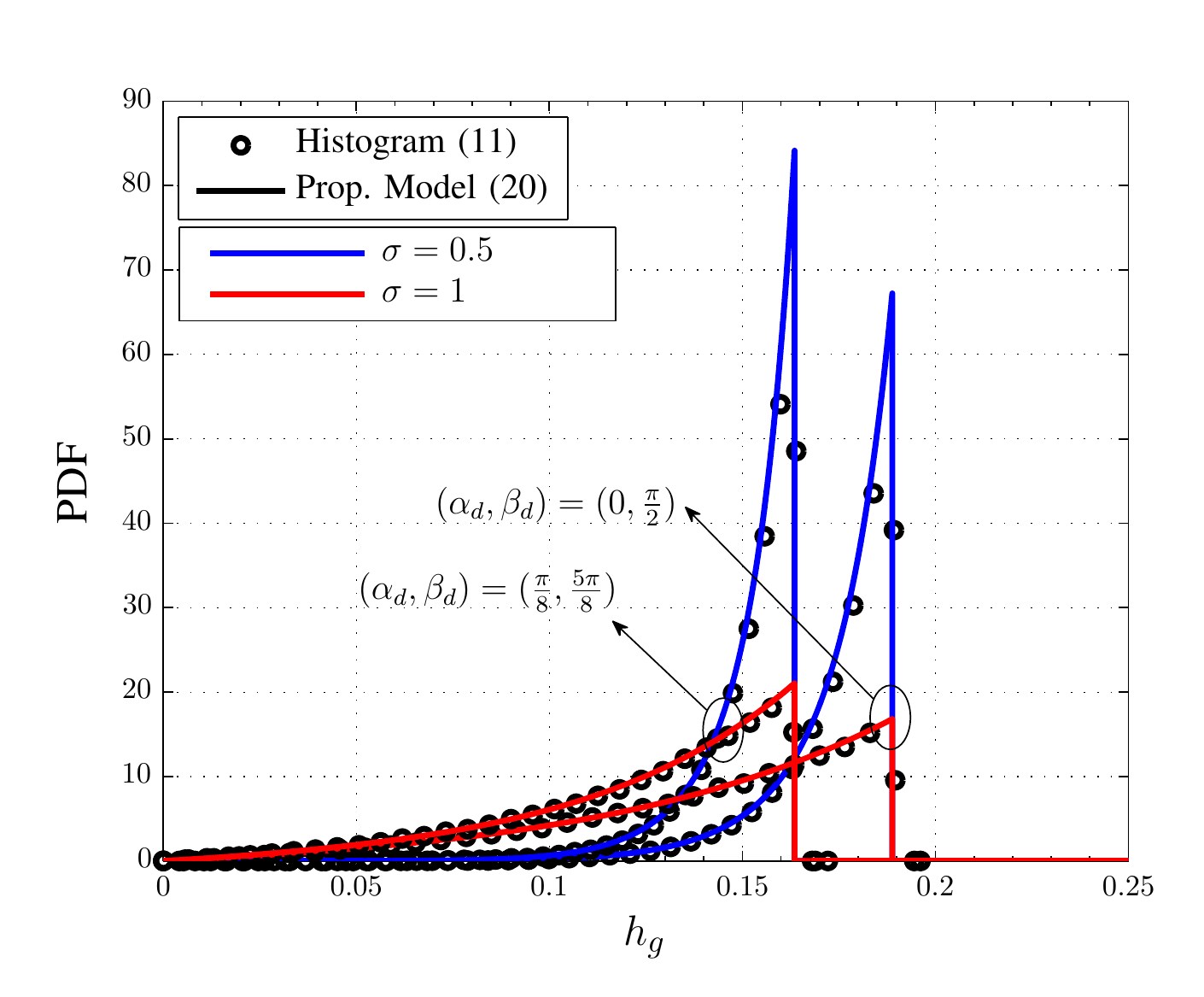}
		\caption{PDF of the GML for independent Gaussian fluctuations.} 
		\label{Fig:PDF_s}
	\end{minipage}
	\hfill
	\begin{minipage}[b]{0.01\textwidth}
	\end{minipage}
	\hfill
	\begin{minipage}[b]{0.49\textwidth}
		\centering
		\hspace{-0.5cm}
		\includegraphics[width=1\linewidth]{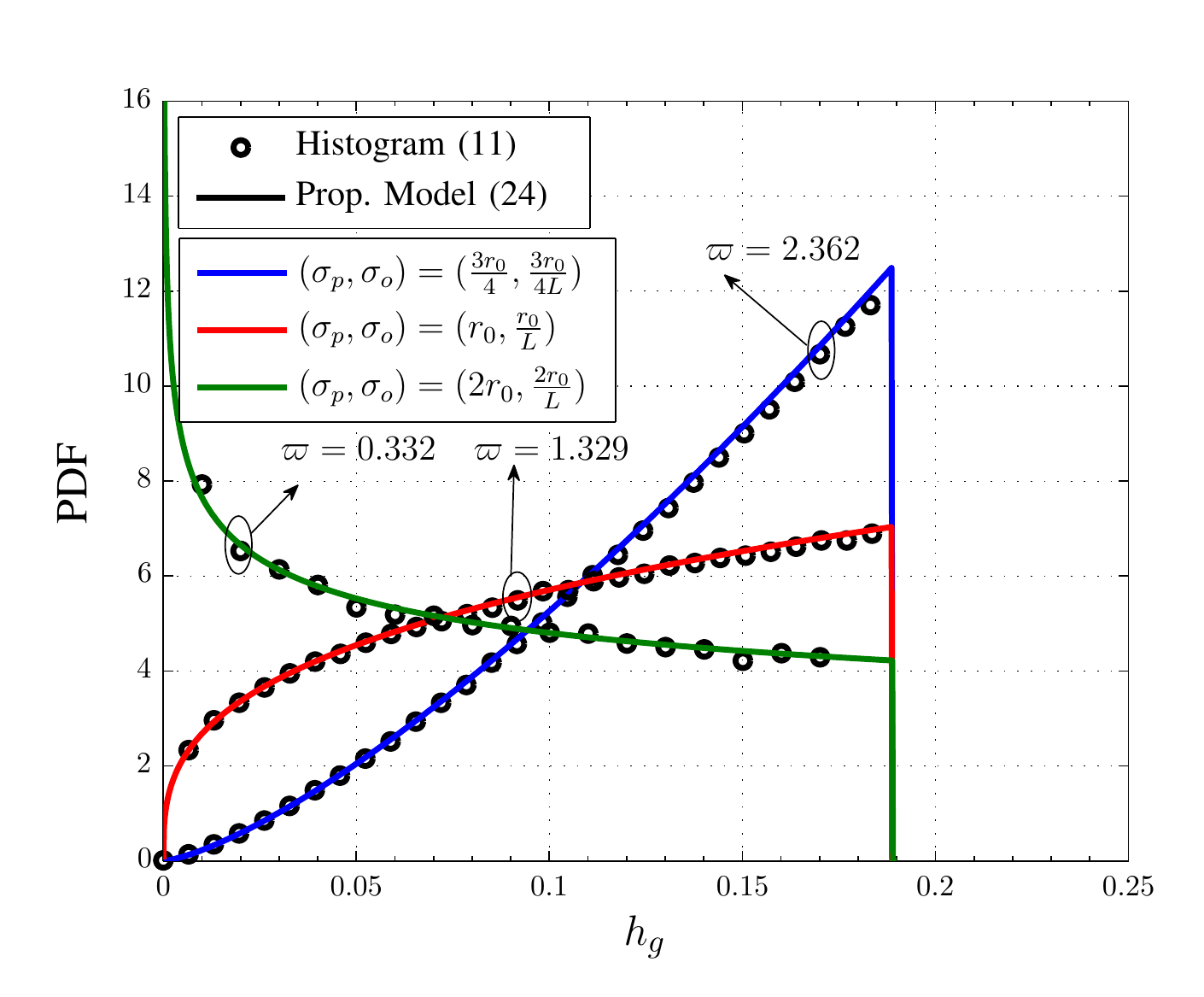}
		\caption{PDF of the GML for independent Gaussian fluctuations and an orthogonal beam w.r.t. the lens plane.}
		\label{Fig:PDF_Increasing}
	\end{minipage}
	\hfill
	\begin{minipage}[b]{0.00\textwidth}
	\end{minipage} 
\end{figure*}

\begin{figure*}[!tbp]
	\centering
	\begin{minipage}[b]{0.49\textwidth}
		\centering
        \includegraphics[width=1\linewidth]{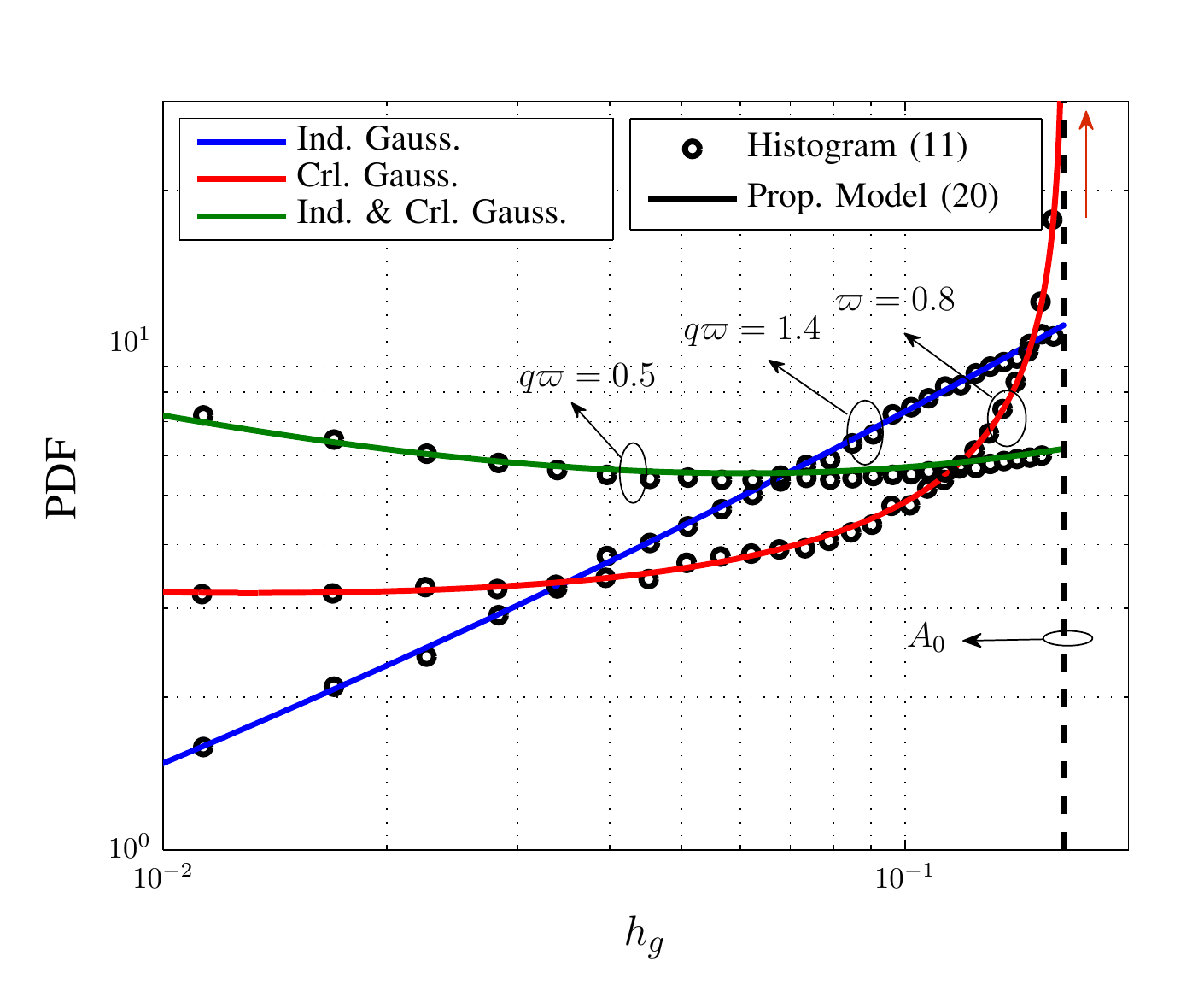}		
		\caption{PDF of the GML for independent and correlated Gaussian fluctuations for  $\zeta=2r_0$ and $(\alpha_d,\beta_d)=(\frac{\pi}{8},\frac{5\pi}{8})$.} 
		\label{Fig:PDF_Breeze}
	\end{minipage}
	\hfill
	\begin{minipage}[b]{0.01\textwidth}
	\end{minipage}
	\hfill
	\begin{minipage}[b]{0.49\textwidth}
		\centering
		\includegraphics[width=1\linewidth]{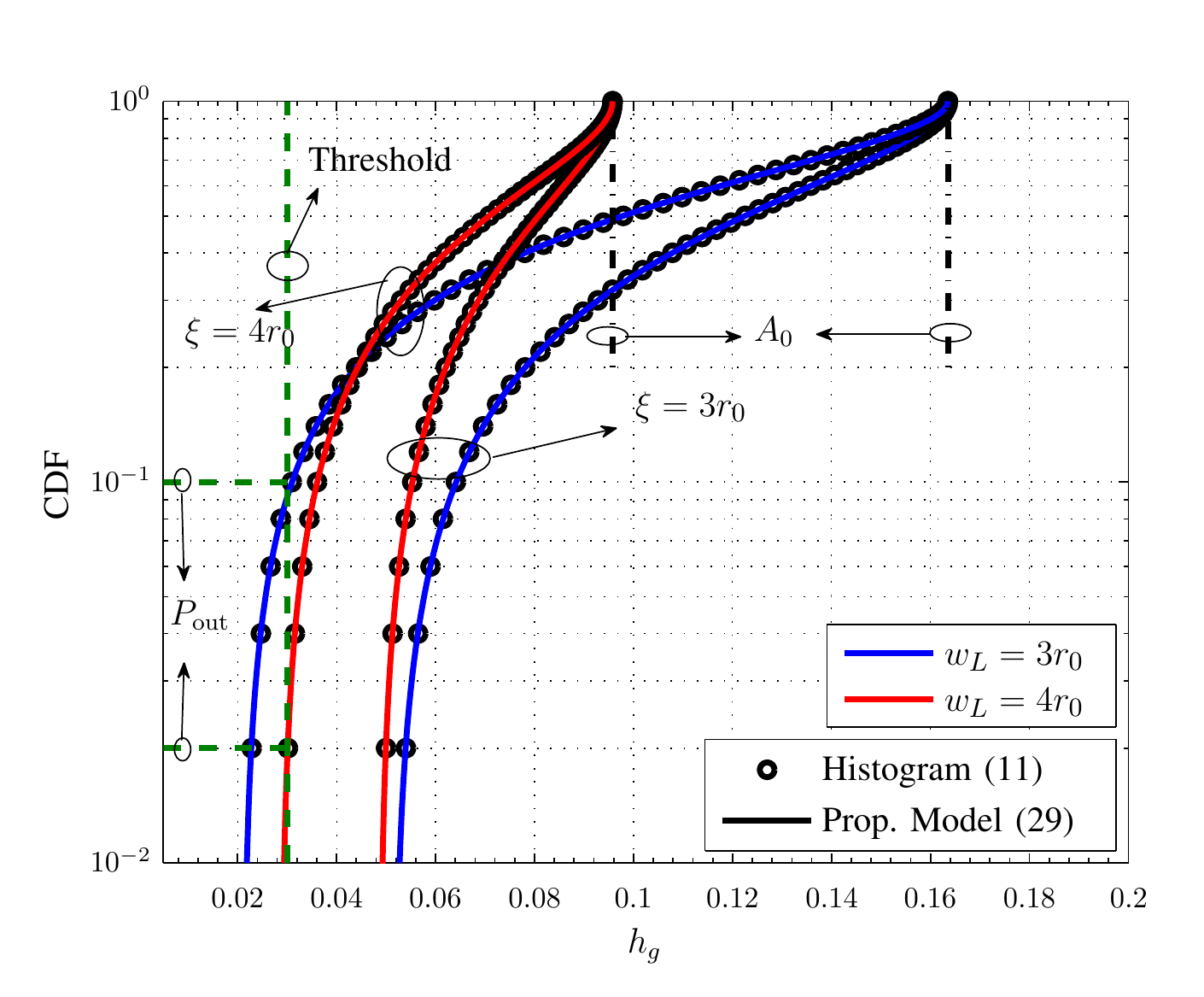}
		\caption{ CDF of the GML for correlated uniform fluctuations with $\xi\in\{3r_0,4r_0\}$ and $(\alpha_d,\beta_d)=(\frac{\pi}{8},\frac{5\pi}{8})$.} 
		\label{Fig:CDF_Unif}
	\end{minipage}
	\hfill
	\begin{minipage}[b]{0.01\textwidth}
	\end{minipage}
\end{figure*}

\begin{figure*}[!tbp]
	\centering
	\begin{minipage}[b]{0.49\textwidth}
		\centering
		\includegraphics[width=1\linewidth]{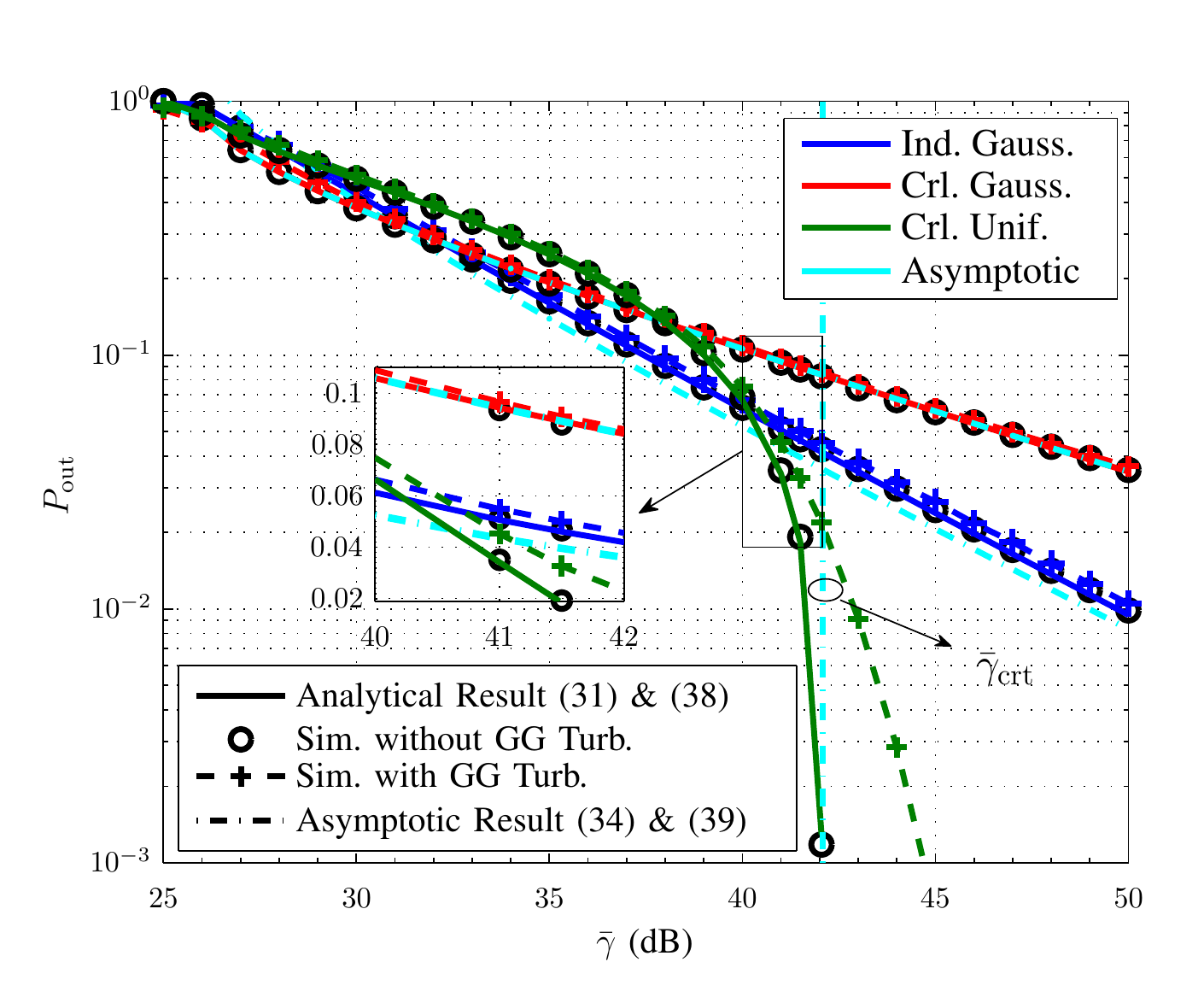}
		\caption{Outage probability vs. $\bar{\gamma}$ for  $\zeta=\xi=r_0$ and $(\alpha_d,\beta_d)$ $=(\frac{\pi}{8},\frac{5\pi}{8})$.}
		\label{Fig:Outage}
	\end{minipage}
	\hfill
	\begin{minipage}[b]{0.01\textwidth}
	\end{minipage}
	\hfill
	\begin{minipage}[b]{0.49\textwidth}
		\centering
		\includegraphics[width=1\linewidth]{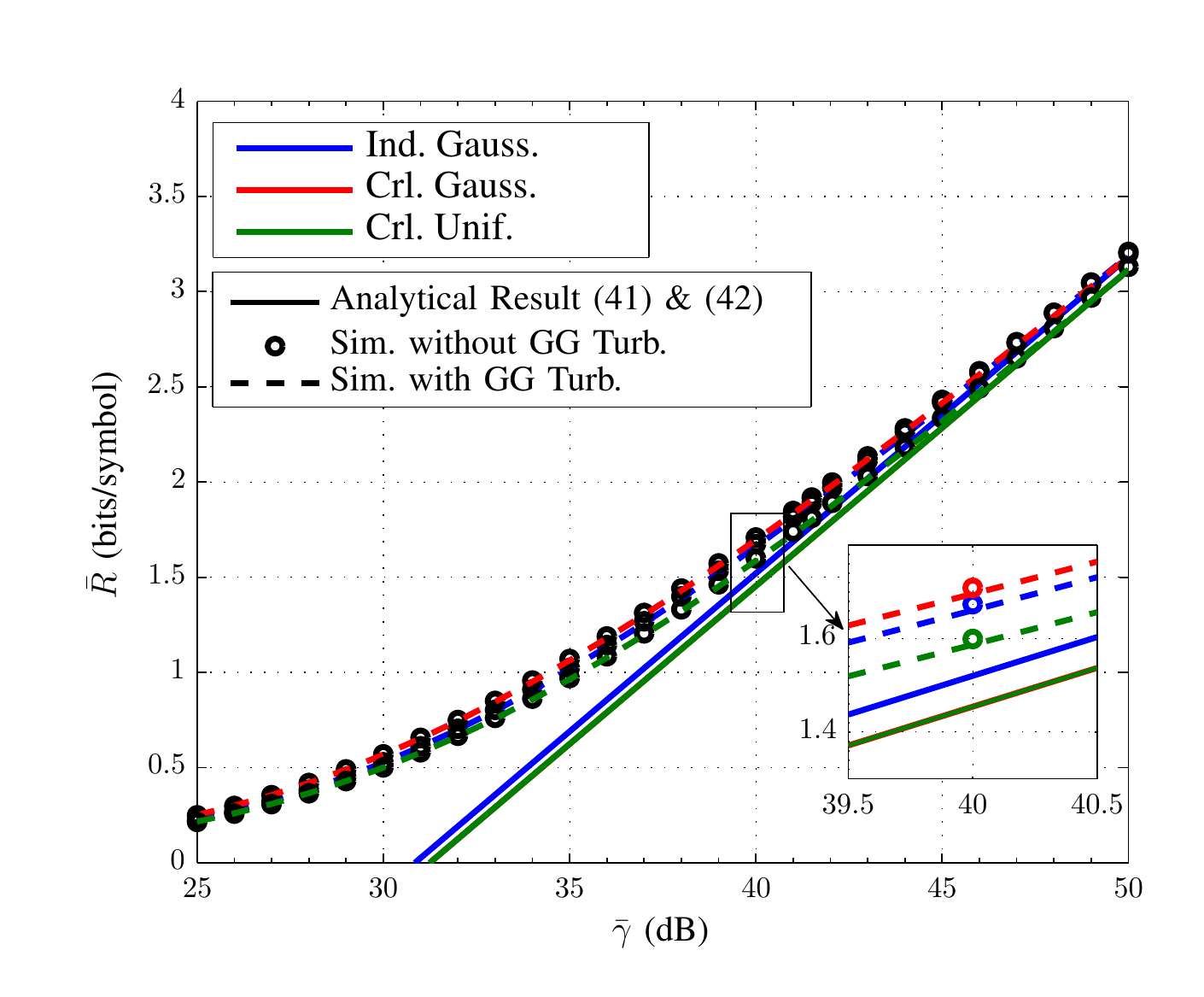}
		\caption{Ergodic rate vs. $\bar{\gamma}$ for  $\zeta=\xi=r_0$ and $(\alpha_d,\beta_d)$ $=(\frac{\pi}{8},\frac{5\pi}{8})$ }
		\label{Fig:Rate} 
	\end{minipage}
	\hfill
	\begin{minipage}[b]{0.01\textwidth}
	\end{minipage} 
\end{figure*}

 In order to quantify the non-orthogonality of the beam w.r.t. the lens plane, we express the mean position of the UAV, $\boldsymbol{\mu}_{\mathbf{r}}$, in spherical coordinates as $(L,\alpha_d,\beta_d)$, i.e., $r_x=L\sin\beta_d\cos\alpha_d$,  $r_y=L\sin\beta_d\sin\alpha_d$, and $r_z = L\cos\beta_d$.  Recall that for a given $\boldsymbol{\mu}_{\mathbf{r}}$, the $\boldsymbol{\mu}_{\boldsymbol{\omega}}$ can be obtained from (\ref{Eq:Angles}). Unless stated otherwise, the default values of the parameters used for the simulations are: $(\alpha_d,\beta_d)=(\frac{\pi}{8},\frac{5\pi}{8})$, $L=500$~m, $h_d=120$~m, $\lambda=1550$~nm, $r_0=10$~cm, $w_L= 30$~cm, $\mathbf{v}=\frac{(3,1,2)}{\|(3,1,2)\|}=(0.8,0.27,0.53)$, and $\boldsymbol{\tau}=\frac{1}{L}\frac{(1,2)}{\|(1,2)\|}=\frac{1}{L}(0.44,0.9)$~\cite{Steve_pointing_error,Alouini_Pointing}. Moreover, the simulation results reported in Figs.~\ref{Fig:PDF_s}-\ref{Fig:Rate} were obtained based on Monte Carlo simulations and $10^6$ realizations of RVs $\mathbf{r}$ and $\boldsymbol{\omega}$. These simulation results are used to verify the accuracy of the assumptions made throughout the paper.

\begin{figure*}
\centering
\includegraphics[width=0.6\linewidth]{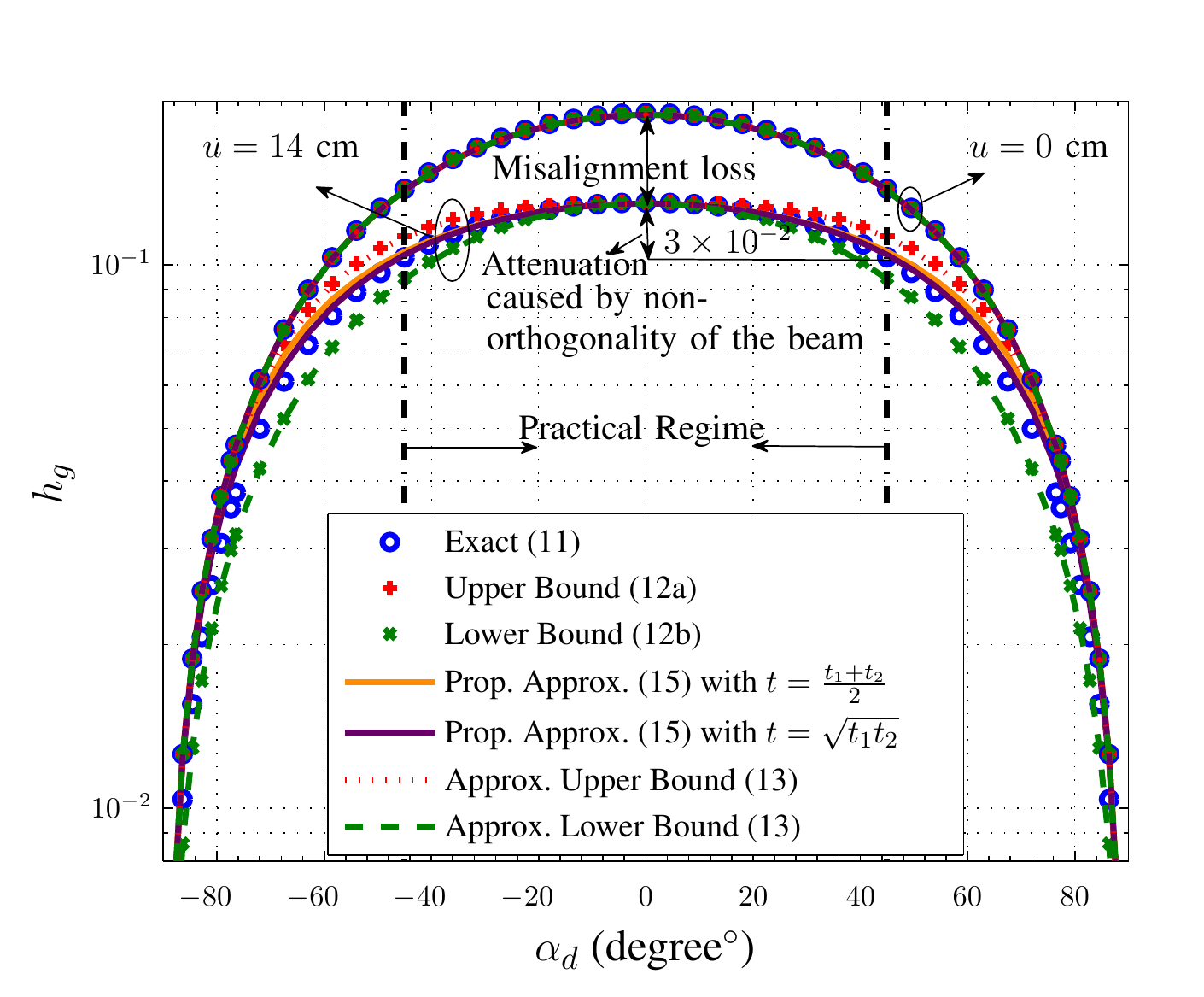}
		\caption{Conditional GML vs. $\alpha_d$ for $\beta_d=\pi/2$.} 
		\label{Fig:Bound}
\end{figure*}
First, we study the impact of non-orthogonality of the beam on the conditional GML and investigate the accuracy of the bounds proposed in Theorem~\ref{Theo:Power}, their corresponding approximations in (\ref{Eq:upper_lower_cal}), and the proposed approximation for the GML in (\ref{Eq:hg_mean}). For Fig.~\ref{Fig:Bound}, the UAV is located in the $x-y$ plane ($\beta_d=\pi/2$) at distance $L$ from the receiver lens and its position on the perimeter of a semicircle with radius $L$ is varied via angle $\alpha_d$ w.r.t. the $x$ axis. In particular, in this figure, we show the conditional GML $h_g$ vs. $\alpha_d$ for two cases of misalignment vector $\mathbf{b}$, namely \textit{i)} $\mathbf{b}=(0,0,0)$~cm which implies that there is no misalignment, i.e., $u=0$~cm, and \textit{ii)} $\mathbf{b}=(0,10,10)$~cm which implies that the center of the beam footprint is outside the receiver lens, i.e., $u\approx 14$~cm$>r_0$. The curve for $u=0$~cm shows the maximum fraction of power, $A_0$, that is collected by the receiver lens for different values of $\alpha_d$, and therefore, the bounds in \eqref{Eq:upper_lower_cal} and the expression proposed for the conditional GML in \eqref{Eq:hg_mean} become identical. Non-zero misalignment causes an additional attenuation $\exp\big(\frac{-2u^2}{tw_L^2}\big)$ to $A_0$, cf. \eqref{Eq:hg_mean}. In Fig.~\ref{Fig:Bound}, this attenuation caused by $14$~cm misalignment is the gap between the curves for $u=0$ and $u=14$~cm at any given $\alpha_d$.
Moreover, the differences between the maximum value of each curve and any other point on that curve show the loss caused by  non-orthogonality of the beam w.r.t. the lens plane. Furthermore, as the beam becomes more non-orthogonal w.r.t. the lens plane, i.e., as $|\alpha_d|$ increases, the channel coefficient $h_g$ decreases and approaches zero when the beam is parallel to the lens, i.e., $|\alpha_d|=\pi/2$. We note that for the practical operating regime of the hovering UAV, $|\alpha_d| \leq \pi/4$ holds. Interestingly, in this regime, the proposed approximations for $h_g$ in (\ref{Eq:hg_mean}), using either the arithmetic or geometric means for $t$, are in good agreement with the simulation results. Besides, in this regime, the loss due to non-orthogonality is small (e.g., $3\times 10^{-2}\approx -1.5$~dB for $u=14$~cm).

Next, in Figs.~\ref{Fig:PDF_s}-\ref{Fig:CDF_Unif}, we study the effect of the random fluctuations of the position and orientation of the UAV on the GML and investigate the accuracy of the statistical models developed for different fluctuation scenarios. Specifically, in Fig.~\ref{Fig:PDF_s}, the PDF of the GML is plotted assuming independent Gaussian fluctuations of the position and orientation of the UAV with standard deviations (SDs) $(\sigma_x,\sigma_y,\sigma_z)=\sigma r_0(0.8,0.27,0.53)$ and $(\sigma_{\theta},\sigma_{\phi})=\frac{\sigma r_0}{L}(0.44,0.9)$. Thereby, $\sigma$ controls the SDs and $\frac{1}{L}$ normalizes the orientation fluctuations w.r.t. the distance. In this figure, we plot the PDF of $h_g$ for $\sigma=0.5$ and $\sigma=1$ and orthogonal ($(\alpha_d,\beta_d)=(0,\frac{\pi}{2})$) and non-orthogonal ($(\alpha_d,\beta_d)=(\frac{\pi}{8},\frac{5\pi}{8})$) beams w.r.t. the lens plane. As can be observed from Fig.~\ref{Fig:PDF_s}, the analytical statistical model proposed in (\ref{Eq:PDF_h_In}) is in perfect agreement with the histogram obtained based on (\ref{Eq:PowerPhotoDetector}). This agreement also validates our assumption in Section IV.B that the main cause of randomness in the GML is the misalignment $u$ and variables $A_0$ and $t$ are practically constant compared to $u$. Moreover, the PDFs for the orthogonal beam, i.e., $(\alpha_d,\beta_d)=(0,\frac{\pi}{2})$, assume non-zero values at larger $h_g(\mathbf{r},\boldsymbol{\omega})$ compared to those for the non-orthogonal beam, i.e.,  $(\alpha_d,\beta_d)=(\frac{\pi}{8},\frac{5\pi}{8})$, since an additional attenuation is caused by the non-orthogonality of the beam. In addition, for larger $\sigma$, the UAV becomes less stable and the probability of smaller channel coefficients increases. Hence, the corresponding PDFs become more heavy tailed.

 Fig.~\ref{Fig:PDF_Increasing} shows the PDF of the GML for the special case considered in Corollary~\ref{Corol:uRayl} and (\ref{Eq:PDF_Rayleigh}), where the beam is orthogonal to the lens plane and $\sigma_x=\sigma_y=\sigma_z=\sigma_p$ and $\sigma_{\theta}=\sigma_{\phi}=\sigma_o$ hold, for different variances of the position and orientation fluctuations, i.e., $\sigma_p=L\sigma_o=r_0(\frac{3}{4},1,2)$. It is observed from Fig.~\ref{Fig:PDF_Increasing} that by increasing $\sigma_p$ and $\sigma_o$, the probability of smaller channel coefficients increases and the PDF becomes more heavy tailed. More specifically, by increasing $\sigma_p$ and $\sigma_o$, the value of $\varpi$ decreases (see the values of $\varpi$ for different $\sigma_p$ and $\sigma_o$ in the figure) and for $\varpi<1$,  $\lim_{h\to 0} f_{h_g}(h)=\infty$ occurs which is consistent with our analytical results in (\ref{Eq:PDF_Rayleigh}).

In Fig.~\ref{Fig:PDF_Breeze}, we investigate the case of correlated Gaussian fluctuations.  To study the impact of correlation, we use identical variances for the positions (orientations), i.e., the main diagonal entries of $\mathbf{Q}_{\mathbf{r}}^{\mathrm{IG}}$ and  $\mathbf{Q}_{\mathbf{r}}^{\mathrm{CG}}$ ($\mathbf{Q}_{\boldsymbol{\omega}}^{\mathrm{IG}}$ and  $\mathbf{Q}_{\boldsymbol{\omega}}^{\mathrm{CG}}$), for the independent and correlated Gaussian scenarios are identical. In other words, we set $(\sigma_x^2,\sigma_y^2,\sigma_z^2)=\zeta^2(v_x^2,v_y^2,v_z^2)$ and 
$(\sigma_{\theta}^2,\sigma_{\phi}^2)=\zeta^2(\tau_{\theta}^2,\tau_{\phi}^2)$. In Fig.~\ref{Fig:PDF_Breeze}, we plot the PDF of the GML for $\zeta=2r_0$, $\mathbf{v}=\frac{(3,4,5)}{\|(3,4,5)\|}=(0.42,0.56,0.7)$, $\boldsymbol{\tau}=(0,0)$, and a non-orthogonal beam w.r.t. the lens plane. It is observed from this figure that for correlated Gaussian fluctuations (Crl. Gauss.), where only the effect of wind is considered, the probabilities of both small and large values for channel coefficient $h_g$ are higher compared to the case when the fluctuations are independent (Ind. Gauss.). In particular, the PDF assumes large values at $A_0$, which is expected based on our analytical results, cf. \eqref{Eq:PDF_halfnormal}, and is also large for small values of $h_g$ since $\varpi<1$ holds for the set of parameters adopted for this figure (see the value of $\varpi$ in the figure). Furthermore, combined independent and correlated fluctuations (Ind. \& Crl. Gauss.) cause the PDF to have larger values for smaller channel coefficients, i.e., it becomes more heavy tailed. Particularly, for the set of parameters adopted in this figure, the PDF for combined independent and correlated Gaussian fluctuations becomes very large at values close to zero since  $q\varpi<1$ holds (see the value of $q\varpi$ in the figure), cf. (\ref{Eq:PDFat0}). 

In Fig.~\ref{Fig:CDF_Unif}, we show the CDF of the GML for correlated uniformly distributed fluctuations for different beam widths, $w_L\in\{3r_0,4r_0\}$, and $\xi$, $\xi\in\{3r_0,4r_0\}$. We observe that the simulation and analytical results are in perfect agreement. Moreover, unlike for Gaussian fluctuations, for uniformly distributed fluctuations, the probability of channel coefficients that are smaller than a certain value, $h_g\leq h_1$, is zero, which is expected based on the analytical expression for the PDF in (\ref{Eq:PDF_h_Unif}). Comparing the curve for $\xi=3r_0$ and the respective curve for $\xi=4r_0$ shows that for the larger $\xi$, since the UAV becomes less stable, the channel quality deteriorates, i.e., the value of $h_1$ for the CDF for $\xi=4r_0$ is smaller than that for $\xi=3r_0$. 
Comparing the curves for different beam widths and a given $\xi$ reveals that, for the wider beam, $w_L=4r_0$, the maximum fraction of power that is collected at the PD, $A_0$, is smaller than $A_0$ for $w_L=3r_0$.
 On the other hand, given a threshold, the wider the beam is, the smaller the outage probability becomes. For instance, for $\xi=4r_0$ in Fig.~\ref{Fig:CDF_Unif}, assuming that an outage occurs when $h_g\leq 0.03$,  we have $A_0=0.16$ and $P_{\mathrm{out}}=0.1$ for $w_L=3r_0$ and $A_0=0.1$ and $P_{\mathrm{out}}=0.02$ for $w_L=4r_0$.  This observation illustrates the trade-off between $A_0$ and $P_{\mathrm{out}}$ for different beam widths.

In Figs.~\ref{Fig:Outage} and \ref{Fig:Rate}, we study the performance of a single UAV-based FSO link in terms of its outage probability and ergodic rate for clear weather conditions, i.e., $\kappa=0.43\times10^{-3}$~m$^{-1}$. In particular, for  Figs.~\ref{Fig:Outage} and \ref{Fig:Rate}, we assume a non-orthogonal beam w.r.t. the lens plane and independent Gaussian, correlated Gaussian, and correlated uniformly distributed fluctuations for the position and orientation of the UAV. For the independent Gaussian fluctuations, we adopt the same covariance matrices for the position and orientation fluctuations as for Fig.~\ref{Fig:PDF_Breeze}. Moreover, we set $\zeta=\xi$ to $r_0$ and $\gamma_{\mathrm{thr}}$ is given by $\gamma_{\mathrm{thr}}=\frac{2\pi}{e}2^{2R_{\mathrm{thr}}-1}$, where $R_{\mathrm{thr}}$ is the transmission rate. In Fig.~\ref{Fig:Outage}, we depict the outage probability vs. SNR ($\bar{\gamma}$) assuming $R_{\mathrm{thr}}=0.5$ bit/symbol in the presence and absence of GG distributed turbulence. We observe that simulation results and analytical results are in perfect agreement for all considered fluctuation models. Furthermore, the gap between the curves with and without GG turbulence is negligible for Gaussian fluctuations which was expected for link lengths, $L$, on the order of several hundred meters, see also the zoomed out part of the figure. The gap is also small for uniform fluctuations for small and medium SNRs but it becomes larger for high SNRs since in this case, for $\bar{\gamma}>\bar{\gamma}_{\mathrm{crt}}$, GG turbulence is the only fading left and the GML is not present anymore, cf. \eqref{Eq:Unif_outage}. Finally, Fig.~\ref{Fig:Outage} confirms the accuracy of the asymptotic outage expression given in (\ref{Eq:Corollary_outage}) for Gaussian fluctuations at high SNRs. 

In  Fig.~\ref{Fig:Rate}, we plot the ergodic rate vs. SNR ($\bar{\gamma}$) for the same fluctuation models as considered in Fig.~\ref{Fig:Outage}. First, the difference between the ergodic rates obtained from simulation with and without GG fading is very small, see also the zoomed out part of the figure. This confirms that the impact of GG turbulence on UAV-based FSO links with lengths on the order of several hundred meters is negligible. Moreover, it is observed that the simulated ergodic rates approach the analytical asymptotic ergodic rates at high SNR for all three fluctuation scenarios. Furthermore, since the asymptotic ergodic rates for different fluctuation models differ only in their constant rate losses $\Delta\bar{R}_g$, cf. (\ref{Eq:R_Geo_highSNR}), they approach one another at high SNR. For instance, for the considered set of parameters, the rate losses for independent Gaussian, correlated Gaussian, and correlated uniform fluctuations are given by $\Delta\bar{R}_g\in\{0.83,0.9,0.9\}$~bits/symbol, respectively. 
Moreover, since the $\Delta\bar{R}_g$ for the correlated Gaussian and correlated uniform scenarios are identical (since $\zeta=\xi$, cf. \eqref{Eq:deltaR}), the respective ergodic rates are equal.

\section{Conclusions}

In this paper, we derived novel statistical models for the FSO fronthaul channel of UAV-based communication systems, by taking into account the non-orthogonality of the laser beam and the random fluctuations of the position and orientation of the UAV. We first modeled the GML as a function of a given position and orientation of the UAV and derived a conditional model. Next, we developed statistical models for the GML assuming independent Gaussian, correlated Gaussian, and correlated uniformly distributed fluctuations for the position and orientation of the UAV which may reflect calm, weakly windy, and strongly windy weather conditions, respectively. Based on the aforementioned channel models, we further analyzed the performance of the UAV-based FSO link in terms of its outage probability and ergodic rate and derived corresponding asymptotic expressions for the high SNR regime. Simulation results validated the presented analysis and revealed important insights for system design. For example, for correlated fluctuations, the probabilities of both small and large values of the channel coefficient are larger than for independent fluctuations. This characteristic leads to a higher outage probability for correlated fluctuations compared to independent fluctuations. Although the specific requirements on the stability of the UAV depend on the targeted application, our simulation results suggested that for high quality FSO links, the standard deviations of the residual fluctuations of the position and orientation of the UAV should not exceed a few $r_0$ ($\approx$ tens of cm) and a few $r_0/L$ ($\approx$ below one mrad), respectively.  We note that the requirements on the permitable fluctuations can be relaxed at the cost of higher power consumption if a wider beam is employed. In particular, when the variance of the fluctuations is relatively large, e.g., due to windy weather conditions or a UAV with low stability, a wider beam is preferable to avoid outages although this causes the average (and the maximum) collected power to decrease. On the other hand, when the variance of the fluctuations is small,  a narrower beam is preferable since it increases the amount of power collected by the PD.

\appendices
\section{}\label{App:Lem_PowerDensity}

$I(y,z)\mathrm{d}y\mathrm{d}z$ is the fraction of power collected in the infinitesimally small area $\mathrm{d}y\mathrm{d}z$, i.e., $\mathrm{d}y\to 0$ and $\mathrm{d}z \to 0$, around the point $(0,y,z)$. Moreover, we use the fact that any point  $(0,y,z)$ in the lens plane is also located in another plane which is perpendicular to the beam line. Therefore, power $I(y,z)\mathrm{d}y\mathrm{d}z$ can be obtained as $I(y,z)\mathrm{d}y\mathrm{d}z=I^{\mathrm{orth}}(l;L)\sin\psi\mathrm{d}y\mathrm{d}z$, where $I^{\mathrm{orth}}(l;L)$ is given in \eqref{Eq:PowerOrthogonal} and $\psi$ is the angle between the beam line and the lens plane which is found from the inner product of the beam direction and a vector orthonormal to the lens plane, i.e., $(1,0,0)$ in the $x$ direction, as follows
\begin{IEEEeqnarray}{lll}\label{Eq:psi}
\sin(\psi)=\frac{\|(1,0,0)\cdot\mathbf{d}\|}{\|(1,0,0)\|\|\mathbf{d}\|}=d_x=\sin\phi \cos\theta.
\end{IEEEeqnarray}
Here, $\mathbf{d}$ is the direction of the beam given in (\ref{Eq:d_angle}) and we exploited $\Vert \mathbf{d}\Vert=1$. Next, we find distances $L$ and $l$. In fact,
$l$ is the distance between point $(0,y,z)$ and the beam line in (\ref{Eq:d_angle}). In general, the distance between a point, $\mathbf{p}$, and a line specified by direction vector $\mathbf{u}$ and a given point, $\mathbf{q}$, on the line can be obtained as $l = \frac{\Vert (\mathbf{p}-\mathbf{q})\times \mathbf{u}\Vert}{\Vert \mathbf{u}\Vert }$.
 For the problem at hand, we choose $\mathbf{p}=(0,y,z)$, $\mathbf{u}=\mathbf{d}$, and $\mathbf{q}=\mathbf{b}$, which leads to
 \iftoggle{OneColumn}{%
\begin{IEEEeqnarray}{lll} \label{Eq:Distance}
l = 
\big\Vert\left(\tilde{y}\cos\phi-\tilde{z}\sin\phi\sin\theta,
\tilde{z}\sin\phi\cos\theta,\tilde{y}\sin{\phi}\cos\theta\right)\big\Vert
=\sqrt{\rho_y\tilde{y}^2+\rho_z\tilde{z}^2+2\rho_{yz}\tilde{y}\tilde{z}},\quad
\end{IEEEeqnarray}
}{%
\begin{IEEEeqnarray}{lll} \label{Eq:Distance}
l = \big\Vert&\big(\tilde{y}\cos\phi-\tilde{z}\sin\phi\sin\theta,
\tilde{z}\sin\phi\cos\theta,\tilde{y}\sin{\phi}\cos\theta\big)\big\Vert
\nonumber\\
&=\sqrt{\rho_y\tilde{y}^2+\rho_z\tilde{z}^2+2\rho_{yz}\tilde{y}\tilde{z}},\quad
\end{IEEEeqnarray}
}
where we exploited $\Vert \mathbf{d}\Vert=1$, replaced $\mathbf{d}$ with (\ref{Eq:BeamLine}), introduced $\tilde{y} = y-b_y$ and $\tilde{z} =z-b_z$, and used $\rho_y$, $\rho_z$, and $\rho_{yz}$ given in Lemma~\ref{Lem:PowerDensity}. Moreover, $L$, the distance between the perpendicular plane w.r.t. the laser beam  that contains point $(0,y,z)$ and the laser source can be bounded as
\begin{IEEEeqnarray}{lll} 
\Vert \mathbf{r} - \mathbf{b} \Vert -  \sqrt{\tilde{y}^2+\tilde{z}^2} \leq L \leq \Vert \mathbf{r} - \mathbf{b} \Vert+ \sqrt{\tilde{y}^2+\tilde{z}^2},
\end{IEEEeqnarray}
where the extreme cases occur if the beam line is parallel to the $y-z$ plane. In particular, we can safely assume that $\Vert \mathbf{r} - \mathbf{b} \Vert \pm  \sqrt{\tilde{y}^2+\tilde{z}^2} \approx \Vert \mathbf{r} \Vert$ holds since the distance between the UAV and the CU, i.e., $\Vert \mathbf{r} \Vert$, is much larger than $\Vert \mathbf{b} \Vert$ and $\sqrt{\tilde{y}^2+\tilde{z}^2}$. Therefore, by substituting $L\approx \Vert \mathbf{r} \Vert$ and (\ref{Eq:Distance}) into (\ref{Eq:PowerOrthogonal}) and using $I(y,z)\mathrm{d}y\mathrm{d}z=I^{\mathrm{orth}}(l;L)\sin\psi\mathrm{d}y\mathrm{d}z$ and (\ref{Eq:psi}), we obtain (\ref{Eq:Intensity}) which completes the proof.

\section{}\label{App:Integral}

In the $y-z$ plane, the contours of power density, $I(y,z)=\bar{I}$, form ellipsoids~given~by 
\iftoggle{OneColumn}{%
\begin{IEEEeqnarray}{lll} \label{Eq:ellipse}
\rho_y(y-b_y)^2+2\rho_{yz}(y-b_y)(z-b_z)+\rho_z(z-b_z)^2=\frac{w^2_L}{2}\ln\left(\frac{2\sin\psi}{\pi w^2_L\bar{I}}\right).\qquad
\end{IEEEeqnarray}
}{%
\begin{IEEEeqnarray}{lll} \label{Eq:ellipse}
\rho_y(y-b_y)^2+2\rho_{yz}(y-b_y)&(z-b_z)+\rho_z(z-b_z)^2\nonumber\\
&=\frac{w^2_L}{2}\ln\left(\frac{2\sin\psi}{\pi w^2_L\bar{I}}\right).\qquad
\end{IEEEeqnarray}
}
 These ellipsoids are centered at point $(b_y,b_z)$ and rotated by angle $\frac{1}{2}\tan^{-1}$ $\big(\frac{2 \rho_{yz}}{\rho_{y} -\rho_{z}}\big)$ counterclockwise. They have minor and major axis lengths of $2\sqrt{{\rho_{\min}}{d}}$ and $2\sqrt{{\rho_{\max}}{d}}$, respectively, where 
 \iftoggle{OneColumn}{%
\begin{IEEEeqnarray}{lll}
 \rho_{\min}=\frac{2}{\rho_y+\rho_z+\sqrt{(\rho_y-\rho_z)^2+4\rho_{zy}^2}},\quad \rho_{\max}=\frac{2}{\rho_y+\rho_z-\sqrt{(\rho_y-\rho_z)^2+4\rho_{zy}^2}}. 
 \end{IEEEeqnarray}
}{%
\begin{IEEEeqnarray}{lll}
 \rho_{\min}=\frac{2}{\rho_y+\rho_z+\sqrt{(\rho_y-\rho_z)^2+4\rho_{zy}^2}},\nonumber\\ \rho_{\max}=\frac{2}{\rho_y+\rho_z-\sqrt{(\rho_y-\rho_z)^2+4\rho_{zy}^2}}. 
 \end{IEEEeqnarray}
}
 $\rho_{\min}$ and $\rho_{\max}$ can be further simplified using the definition of $\rho_y$, $\rho_z$, and $\rho_{yz}$ in Lemma~\ref{Lem:PowerDensity} as $\rho_{\min}=1$ and $\rho_{\max}=\frac{1}{\sin^2\phi\cos^2\theta}$.

In order to obtain the lower and upper bounds for $h_g(\mathbf{r},\boldsymbol{\omega})$ in Theorem~\ref{Theo:Power}, we substitute the contour in (\ref{Eq:ellipse}) by two rotated elliptic contours which have the same axis lengths $\rho_{\min}=1$ and $\rho_{\max}=\frac{1}{\sin^2\phi\cos^2\theta}$; however, their main axes are either perpendicular or parallel to the line connecting $(b_y,b_z)$ and the origin, respectively, see Fig.~\ref{Fig:Contour}. Moreover, without loss of generality, we can define a new coordinate system by rotating the $y$ and $z$ axes by angle $\tan^{-1}(\frac{b_z}{b_y})$ such that the center of the ellipsoid in (\ref{Eq:ellipse}) lies on the rotated $y$ axis, i.e., the center becomes $(u,0)$ in the new coordinate system, where $u=\sqrt{b_y^2+b_z^2}$. Note that the circular lens has the same description in the new and the old coordinate systems.  This leads to lower and upper bounds $h_g^{\mathrm{low}}(\mathbf{r},\boldsymbol{\omega})$ and $h_g^{\mathrm{upp}}(\mathbf{r},\boldsymbol{\omega})$, respectively, as given in Theorem~\ref{Theo:Power} and completes the proof.

\section{}\label{App:Low_Upp_approx}
The following integral was approximated in \cite[Appendix]{Steve_pointing_error}
\iftoggle{OneColumn}{%
\begin{IEEEeqnarray}{lll}\label{Eq:Steev_Approx} 
	&\frac{2}{\pi w^2_L}
\underset{(y,z)\in {\mathcal{A}}}{\iint}
	\exp\left(-\frac{2}{ w^2_L}\left((y-u)^2+{z}^2\right)\right) 
	\mathrm{d}y\mathrm{d}z\nonumber\\
	&\overset{(a)}{\approx}\frac{2}{\pi w^2_L}
	\underset{(y,z)\in \bar{\mathcal{A}}}{\iint}
	\exp\left(-\frac{2}{ w^2_L}\left((y-u)^2+{z}^2\right)\right) 
	\mathrm{d}y\mathrm{d}z\overset{(b)}{\approx} A_0\exp\left(\frac{-2u^2}{tw_L^2}\right),
\end{IEEEeqnarray}
}{%
\begin{IEEEeqnarray}{lll}\label{Eq:Steev_Approx} 
	&\frac{2}{\pi w^2_L}
\underset{(y,z)\in {\mathcal{A}}}{\iint}
	\exp\left(-\frac{2}{ w^2_L}\left((y-u)^2+{z}^2\right)\right) 
	\mathrm{d}y\mathrm{d}z\nonumber\\
	&\overset{(a)}{\approx}\frac{2}{\pi w^2_L}
	\underset{(y,z)\in \bar{\mathcal{A}}}{\iint}
	\exp\left(-\frac{2}{ w^2_L}\left((y-u)^2+{z}^2\right)\right) 
	\mathrm{d}y\mathrm{d}z\nonumber\\
	&\overset{(b)}{\approx} A_0\exp\left(\frac{-2u^2}{tw_L^2}\right),
\end{IEEEeqnarray}
}
where $A_0=[\mathrm{erf}(\nu)]^2$, $t=\frac{\sqrt{\pi}\mathrm{erf}(\nu)}{\sqrt{2}\nu\exp(-\nu^2)}$, and $\nu=\frac{\sqrt{\pi}r_0}{\sqrt{2}w_L}$. In \eqref{Eq:Steev_Approx}, equality $(a)$ follows from approximating the circular PD, i.e., $(y,z)\in {\mathcal{A}}$, by a square lens of equal area, i.e., $(y,z)\in \bar{\mathcal{A}}\triangleq \Big\{(y,z)\mid y,z\in\big[-\frac{\sqrt{\pi}r_0}{2}$ $,\frac{\sqrt{\pi}r_0}{2}\big]\Big\}$ and equality $(b)$ is obtained using the Taylor series of function $\exp(\cdot)$. In the following, we use \eqref{Eq:Steev_Approx} to approximate $h_g^{\mathrm{low}}$ and $h_g^{\mathrm{upp}}$ in \eqref{Eq:upper_lower_bound} with $\widetilde{h}_g^{\mathrm{low}}$ and $\widetilde{h}_g^{\mathrm{upp}}$, respectively. We derive $\widetilde{h}_g^{\mathrm{low}}$ since obtaining $\widetilde{h}_g^{\mathrm{upp}}$ follows similar steps.
By approximating $\mathcal{A}$ with $\bar{\mathcal{A}}$ and defining new variable $\hat{z}=\sin\phi\cos\theta z$ for $h_g^{\mathrm{low}}$ in (\ref{Eq:upper_lower_bound}a), $\widetilde{h}_g^{\mathrm{low}}$ is obtained as
\iftoggle{OneColumn}{%
\begin{IEEEeqnarray}{lll}\label{Eq:Generic_low_upp_relaxed}
\widetilde{h}_g^{\mathrm{low}}=\frac{2\sin\psi}{\pi \sin\phi\cos\theta w^2_L}
	\underset{({y},\hat{z})\in \hat{\mathcal{A}}}{\iint}
	\exp\left(-\frac{2}{ w^2_L}\left(({y}-{u})^2+{\hat{z}}^2\right)\right) 
	\mathrm{d}{y}\mathrm{d}\hat{z},
\end{IEEEeqnarray}
}{%
\begin{IEEEeqnarray}{lll}\label{Eq:Generic_low_upp_relaxed}
\widetilde{h}_g^{\mathrm{low}}=\frac{2\sin\psi}{\pi \sin\phi\cos\theta w^2_L}\times\nonumber\\
	\underset{({y},\hat{z})\in \hat{\mathcal{A}}}{\iint}
	\exp\left(-\frac{2}{ w^2_L}\left(({y}-{u})^2+{\hat{z}}^2\right)\right) 
	\mathrm{d}{y}\mathrm{d}\hat{z},
\end{IEEEeqnarray}
}
where $\hat{\mathcal{A}}\triangleq \Big\{({y},\hat{z})\mid {y}\in\big[-\frac{\sqrt{\pi}r_0}{2},\frac{\sqrt{\pi}r_0}{2}\big],\hat{z}\in$ $\big[-\frac{\sqrt{\pi}|\sin\phi\cos\theta|r_0}{2},\frac{\sqrt{\pi}|\sin\phi\cos\theta|r_0}{2}\big]\Big\}$. The integral in \eqref{Eq:Generic_low_upp_relaxed} is similar to the second integral in \eqref{Eq:Steev_Approx} except that $\hat{\mathcal{A}}$ corresponds to a rectangular area whereas $\bar{\mathcal{A}}$ is a square area. Using a similar technique as the one used in \cite[Appendix]{Steve_pointing_error}, we approximate \eqref{Eq:Generic_low_upp_relaxed} as in (\ref{Eq:upper_lower_cal}). This completes the proof.

\section{}\label{App:Theo_PDF_r}

In the following, we first determine the PDF of $u$ and subsequently obtain the PDF of $h_g$ from \eqref{Eq:PDF_h_PDF_u}. To do so, we first simplify the expressions for $b_y$ and $b_z$ in \eqref{Eq:FootPrint_Center} by replacing $\tan\theta$, $\cot\phi$, and $\frac{1}{\cos\theta}$ by their respective Taylor series (assuming $\epsilon_{\theta}=\theta-\mu_{\theta}$ and $\epsilon_{\phi}=\phi-\mu_{\phi}$, cf. (\ref{Eq:r_omega_RV}), are very small) and then relating the PDF of $u$ to that of $b_y$ and $b_z$ exploiting $u=\sqrt{b_y^2+b_z^2}$.
In particular, we obtain
\iftoggle{OneColumn}{%
\begin{IEEEeqnarray}{lll} \label{Eq:fyfz_norm}
\underset{\mathbf{r}\to \boldsymbol{\mu}_{\mathbf{r}},\boldsymbol{\omega}\to\boldsymbol{\mu}_{\boldsymbol{\omega}}}{\lim}  b_y = \epsilon_y+c_1\epsilon_x+c_2\epsilon_{\theta}\qquad\text{and}\qquad
\underset{\mathbf{r}\to \boldsymbol{\mu}_{\mathbf{r}},\boldsymbol{\omega}\to\boldsymbol{\mu}_{\boldsymbol{\omega}}}{\lim} b_z = \epsilon_z+c_3\epsilon_{\phi}+c_4\epsilon_{\theta}+c_5\epsilon_x,
\end{IEEEeqnarray}
}{%
\begin{IEEEeqnarray}{lll} \label{Eq:fyfz_norm}
\underset{\mathbf{r}\to \boldsymbol{\mu}_{\mathbf{r}},\boldsymbol{\omega}\to\boldsymbol{\mu}_{\boldsymbol{\omega}}}{\lim}  b_y = \epsilon_y+c_1\epsilon_x+c_2\epsilon_{\theta}\qquad\text{and}\nonumber\\
\underset{\mathbf{r}\to \boldsymbol{\mu}_{\mathbf{r}},\boldsymbol{\omega}\to\boldsymbol{\mu}_{\boldsymbol{\omega}}}{\lim} b_z = \epsilon_z+c_3\epsilon_{\phi}+c_4\epsilon_{\theta}+c_5\epsilon_x,
\end{IEEEeqnarray}
}
where constants $c_1$-$c_5$ are given in Theorem~\ref{Theo:PDF_r}. To obtain (\ref{Eq:fyfz_norm}), we drop the terms with orders higher than one, e.g., $\epsilon_{\theta}\epsilon_{\phi}$. We note that (\ref{Eq:fyfz_norm}) is valid for all considered fluctuation models. Now, assuming independent Gaussian fluctuations, we add superscript $^\mathrm{IG}$ to $\epsilon_s,\,\,s\in\{x,y,z,\theta,\phi\}$ and $(b_y,b_z)$ in (\ref{Eq:fyfz_norm}), cf. (\ref{Eq:Fluc_Ind_Gauss}). Since $b_y^\mathrm{IG}$ and $b_z^\mathrm{IG}$ are sums of Gaussian RVs, they are Gaussian distributed, too. However, $b_y^\mathrm{IG}$ and $b_z^\mathrm{IG}$ are correlated since $\epsilon_{x}^{\mathrm{IG}}$ and $\epsilon_{\theta}^{\mathrm{IG}}$ appear in the expressions for both. The joint distribution of $b_y^\mathrm{IG}$ and $b_z^\mathrm{IG}$ is a bivariate Gaussian distribution $(b_y^\mathrm{IG},b_z^\mathrm{IG})\triangleq \mathbf{b}_{yz}^\mathrm{IG}\sim\mathcal{N}(\mathbf{0},\boldsymbol{\Sigma}_{\mathrm{IG}})$ where $\boldsymbol{\Sigma}_{\mathrm{IG}}$ is given in (\ref{Eq:Cov}). Let $\boldsymbol{\Sigma}_{\mathrm{IG}}=\mathbf{U}\boldsymbol{\Lambda}\mathbf{U}^\mathsf{T}$ be the eigenvalue decomposition of $\boldsymbol{\Sigma}_{\mathrm{IG}}$ where $\boldsymbol{\Lambda}$ is a diagonal matrix with elements $\lambda_1$ and $\lambda_2$ and $\mathbf{U}$ is a unitary matrix, i.e., $\mathbf{U}^\mathsf{T}\mathbf{U}=\mathbf{I}$. Using these definitions, it is easy to show that $\mathbf{b}_{yz}^\mathrm{IG}\sim\mathbf{g}\mathbf{U}^\mathsf{T}$ where $\mathbf{g}=(g_y,g_z)\sim\mathcal{N}(\mathbf{0},\boldsymbol{\Lambda})$. Now, we can express $u$ in terms of $\mathbf{g}$ as follows
\iftoggle{OneColumn}{%
\begin{IEEEeqnarray}{lll}\label{Eq:u_f} 
u =\sqrt{(b_y^\mathrm{IG})^2+(b_z^\mathrm{IG})^2}=\sqrt{\mathbf{b}_{yz}^\mathrm{IG}(\mathbf{b}^\mathrm{IG}_{yz})^\mathsf{T}} \sim \sqrt{\mathbf{g}\mathbf{U}^\mathsf{T}\mathbf{U} \mathbf{g}^\mathsf{T}}
= \sqrt{g_y^2+g_z^2}. \quad
\end{IEEEeqnarray}
}{%
\begin{IEEEeqnarray}{lll}\label{Eq:u_f} 
u &=\sqrt{(b_y^\mathrm{IG})^2+(b_z^\mathrm{IG})^2}=\sqrt{\mathbf{b}_{yz}^\mathrm{IG}(\mathbf{b}^\mathrm{IG}_{yz})^\mathsf{T}} \nonumber\\
&\sim \sqrt{\mathbf{g}\mathbf{U}^\mathsf{T}\mathbf{U} \mathbf{g}^\mathsf{T}}
= \sqrt{g_y^2+g_z^2}. \quad
\end{IEEEeqnarray}
}
Since $g_y$ and $g_z$ are independent zero-mean Gaussian RVs with non-identical variances,  $u$ follows a Hoyt (Nakagami-q) distribution with PDF $f_u(u)=\frac{1+q^2}{q\Omega}u\exp\left(-\frac{(1+q^2)^2}{4q^2\Omega}u^2\right)I_0\left(\frac{1-q^4}{4q^2\Omega}u^2\right)$, where $q=\sqrt{\frac{\min\{\lambda_1,\lambda_2\}}{\max\{\lambda_1,\lambda_2\}}}$  and $\Omega=\lambda_1+\lambda_2$ \cite{Nakagami_Hoyt,Alouini_Pointing}. Substituting $f_u(u)$ into (\ref{Eq:PDF_h_PDF_u}), the PDF of the GML can be obtained as in (\ref{Eq:PDF_h_In}). This completes the proof.

\section{}\label{App:Theo_PDF_u_Gcorrelated}

For correlated Gaussian fluctuations, we replace $\epsilon_s,\,\,s\in\{x,y,z,\theta,\phi\}$ in (\ref{Eq:fyfz_norm})  with $\epsilon_s^{\mathrm{IG}}+\epsilon_s^{\mathrm{CG}}$ according to the definition in (\ref{Eq:Gaussian_Correlated}). After eliminating the terms of order higher than one, similar to Appendix~\ref{App:Theo_PDF_r}, $(b_y,b_z)$ for the correlated Gaussian scenario, denoted by $(b_y^{\mathrm{CG}},b_z^{\mathrm{CG}})$, is obtained as
\iftoggle{OneColumn}{%
\begin{IEEEeqnarray}{lll} \label{Eq:fyfz_norm2}
	\underset{\underset{\boldsymbol{\omega}\to\boldsymbol{\mu}_{\boldsymbol{\omega}}}{\mathbf{r}\to \boldsymbol{\mu}_{\mathbf{r}}}}{\lim}  b_y^{\mathrm{CG}} = \epsilon_y^{\mathrm{IG}}+c_1\epsilon_x^{\mathrm{IG}}+c_2\epsilon_{\theta}^{\mathrm{IG}}+c_6\delta^{\mathrm{G}}\quad\text{and}\quad
	\underset{\underset{\boldsymbol{\omega}\to\boldsymbol{\mu}_{\boldsymbol{\omega}}}{\mathbf{r}\to \boldsymbol{\mu}_{\mathbf{r}}}}{\lim} b_z^{\mathrm{CG}} = \epsilon_z^{\mathrm{IG}}+c_3\epsilon_{\phi}^{\mathrm{IG}}+c_4\epsilon_{\theta}^{\mathrm{IG}}+c_5\epsilon_x^{\mathrm{IG}}+c_7\delta^{\mathrm{G}}.\qquad
\end{IEEEeqnarray}
}{%
\begin{IEEEeqnarray}{lll} \label{Eq:fyfz_norm2}
	\underset{\underset{\boldsymbol{\omega}\to\boldsymbol{\mu}_{\boldsymbol{\omega}}}{\mathbf{r}\to \boldsymbol{\mu}_{\mathbf{r}}}}{\lim}  b_y^{\mathrm{CG}} = \epsilon_y^{\mathrm{IG}}+c_1\epsilon_x^{\mathrm{IG}}+c_2\epsilon_{\theta}^{\mathrm{IG}}+c_6\delta^{\mathrm{G}}\quad\text{and}\nonumber\\
	\underset{\underset{\boldsymbol{\omega}\to\boldsymbol{\mu}_{\boldsymbol{\omega}}}{\mathbf{r}\to \boldsymbol{\mu}_{\mathbf{r}}}}{\lim} b_z^{\mathrm{CG}} = \epsilon_z^{\mathrm{IG}}+c_3\epsilon_{\phi}^{\mathrm{IG}}+c_4\epsilon_{\theta}^{\mathrm{IG}}+c_5\epsilon_x^{\mathrm{IG}}+c_7\delta^{\mathrm{G}}.\qquad
\end{IEEEeqnarray}
}
Since $b_y^{\mathrm{CG}}$ and $b_z^{\mathrm{CG}}$ are again sums of Gaussian RVs, they are also Gaussian RVs. Therefore, with the same reasoning as in Appendix~\ref{App:Theo_PDF_r}, $u$ follows a Hoyt (Nakagami-q) distribution with parameters $q=\sqrt{\frac{\min\{\lambda_1,\lambda_2\}}{\max\{\lambda_1,\lambda_2\}}}$  and $\Omega=\lambda_1+\lambda_2$. Here, $\lambda_1$ and $\lambda_2$ are now the eigenvalues of $\boldsymbol{\Sigma}_{\mathrm{T}}$ given in Theorem~\ref{Theo:PDF_u_Gcorrelated}. Therefore, formally the same expression for the PDF of the GML is obtained as for independent Gaussian fluctuations. This completes the proof.

\section{}\label{App:Corol_Gauss_Single}
In this case, we obtain $\boldsymbol{\Sigma}_{\mathrm{T}}=\boldsymbol{\Sigma}_{\mathrm{CG}}$ which has one non-zero eigenvalue $\lambda_1=\zeta^2(c_6^2+c_7^2)$, i.e., $q=0$, and  $(b_y^{\mathrm{CG}},b_z^{\mathrm{CG}})=(c_6\delta^{\mathrm{G}},c_7\delta^{\mathrm{G}})$ which results in $u=\sqrt{c_6^2+c_7^2}\,\,|\delta^{\mathrm{G}}|$, cf. \eqref{Eq:fyfz_norm2} in Appendix~\ref{App:Theo_PDF_u_Gcorrelated}. Since $\delta^{\mathrm{G}}$ follows the Gaussian distribution, $u$ follows a single-sided Gaussian distribution with the PDF given by $f_u(u)=\frac{\sqrt{2}}{\sqrt{\pi\lambda_1}}\exp\left(-\frac{u^2}{2\lambda_1}\right)$. Based on the distribution of $u$, the PDF of $f_{h_g}(h)$ in (\ref{Eq:PDF_halfnormal}) is obtained using \eqref{Eq:PDF_h_PDF_u}, which completes the proof.

\section{}\label{App:Theo_PDF_r_Uniform}
We replace $\epsilon_s$ in (\ref{Eq:fyfz_norm}) with $\epsilon_s^{\mathrm{CU}},\,\,s\in\{x,y,z,\theta,\phi\}$, cf. (\ref{Eq:Uniform_r_w}). Therefore, we have 
\begin{IEEEeqnarray}{lll} \label{Eq:fyfz_Uni}
\underset{\mathbf{r}\to \boldsymbol{\mu}_{\mathbf{r}},\boldsymbol{\omega}\to\boldsymbol{\mu}_{\boldsymbol{\omega}}}{\lim}  b_y^{\mathrm{CU}} = c_6\delta^{\mathrm{U}}\,\,\text{and}\,\,
\underset{\mathbf{r}\to \boldsymbol{\mu}_{\mathbf{r}},\boldsymbol{\omega}\to\boldsymbol{\mu}_{\boldsymbol{\omega}}}{\lim} b_z^{\mathrm{CU}} = c_7\delta^{\mathrm{U}},\quad\,\,\,
\end{IEEEeqnarray}
where constants $c_6$ and $c_7$ are given in Theorem~\ref{Theo:PDF_u_Gcorrelated}. Using (\ref{Eq:fyfz_Uni}), $u$ is obtained as
\iftoggle{OneColumn}{%
\begin{IEEEeqnarray}{lll} \label{Eq:U2}
u=\sqrt{(b_y^{\mathrm{CU}})^2+(b_z^{\mathrm{CU}})^2}=\sqrt{c_6^2+c_7^2}\,\,|\delta^{\mathrm{U}}|\sim\mathcal{U}\left(0,\sqrt{3(c_6^2+c_7^2)}\,\xi\right).
\end{IEEEeqnarray}
}{%
\begin{IEEEeqnarray}{lll} \label{Eq:U2}
u&=\sqrt{(b_y^{\mathrm{CU}})^2+(b_z^{\mathrm{CU}})^2}\nonumber\\
&=\sqrt{c_6^2+c_7^2}\,\,|\delta^{\mathrm{U}}|\sim\mathcal{U}\left(0,\sqrt{3(c_6^2+c_7^2)}\,\xi\right).
\end{IEEEeqnarray}
}
Substituting the uniform distribution in (\ref{Eq:U2}) into (\ref{Eq:PDF_h_PDF_u}) leads to (\ref{Eq:PDF_h_Unif}) in Theorem~\ref{Theo:PDF_u2_Uniform}. This completes the proof.

\section{}\label{App:Corollary}

 The symmetric difference of the first-order Marcum Q-function is approximated by $\underset{(a,b)\to \infty}{\lim }Q(a,b)-Q(b,a)= 1-\left(\sqrt{\frac{a}{b}}+\sqrt{\frac{b}{a}}\right)Q\left(a-b\right)$  \cite{MarcumQ}.
 Hence, $P_{\mathrm{out}}$ in (\ref{Eq:CDF_h}) is simplified to $P_{\mathrm{out}}=\left(\sqrt{\frac{a}{b}}+\sqrt{\frac{b}{a}}\right)$ $Q\left(a-b\right)$. Next, we use $Q(x)\approx \frac{e^{-\frac{1}{2}x^2}}{\sqrt{2\pi}x}$ to approximate the Gaussian Q-function at large values, which  leads to  the simplified expression for the outage probability in (\ref{Eq:Corollary_outage}).  This completes the proof.

\section*{Acknowledgements} The authors would like to thank Prof. Mohamed-Slim Alouini and Prof. Steve Hranilovic for helpful and insightful discussions on the uniform distribution of the fluctuations of the position and orientation of the UAV during IEEE ICC 2018.

\bibliographystyle{IEEEtran}
\bibliography{My_Citation_01-05-2019}

\end{document}